\documentclass{aa}  
\usepackage[colorlinks=true,urlcolor=blue,citecolor=blue,linkcolor=blue,breaklinks=true]{hyperref}

\usepackage{graphicx}
\usepackage{txfonts}
\usepackage{algorithm2e}
\usepackage{algpseudocode}
\RestyleAlgo{ruled} 

\newcommand{\diar}[1]{#1}
\newcommand{\diarp}[1]{#1}
\newcommand{\diarpp}[1]{#1}

\begin{document}

   \title{\diar{Jacobian-free Newton-Krylov method for multilevel nonlocal thermal equilibrium radiative transfer problems}}


   \author{D. Arramy
          \and
          J. de la Cruz Rodr\'iguez
          \and
          J. Leenaarts
          }

   \institute{Institute for Solar Physics, Dept. of Astronomy,
              Stockholm University, AlbaNova University Centre, SE-106 91 Stockholm, Sweden \\
              \email{dimitri.arramy@astro.su.se}
             }

   \date{Received ; accepted }

 
  \abstract
   {The calculation of the emerging radiation from a model atmosphere requires knowledge of the emissivity and absorption coefficients, which are proportional to the atomic level population \diar{densities} of the levels involved in each transition. Due to the intricate \diar{inter}dependence of the radiation field and the physical state of the atoms, iterative methods are required in order to \diar{calculate} the atomic \diar{level} population \diar{densities}. A variety of different methods have been proposed to solve this problem, which is known as the \diar{n}onlocal \diar{t}hermodynamical \diar{e}quilibrium (NLTE) problem.}
   {Our goal is to develop an efficient and rapidly converging method to solve the NLTE problem under the assumption of statistical equilibrium. In particular, we explore whether the Jacobian-\diar{F}ree Newton\diar{-Krylov (JFNK)} method can be used. This method does not require an explicit construction of the Jacobian matrix because it estimates the new correction with the Krylov-subspace method.}
   {We implemented an NLTE radiative transfer code with overlapping bound-bound and bound-free transitions. This solved the statistical equilibrium equations using a \diar{JFNK} method, assuming a depth-stratified plane-parallel atmosphere. As a reference, we also implemented the \cite{1992A&A...262..209R} method based on linearization and operator splitting.}
   {Our tests with the Fontenla, Avrett and Loeser C model atmosphere (FAL-C) and two different six-level \ion{Ca}{II} and \ion{H}{I} atoms show that the \diar{JFNK} method can converge faster than our reference case by up to a factor $2$. This number is evaluated in terms of the total number of evaluations of the formal solution of the radiative transfer equation for all frequencies and directions. This method can also reach a lower \diar{residual error} compared to the reference case.}
   {The JFNK method we developed poses a new alternative to solving the NLTE problem. Because it is not based on operator splitting with a local approximate operator, it can improve the convergence of the NLTE problem in highly scattering cases. One major advantage of this method is that it is expected to allow for a direct implementation of more complex problems, such as overlapping transitions from different active atoms, charge conservation, or a more efficient treatment of partial redistribution, without having to explicitly linearize the equations.}


   \maketitle
%

\section{Introduction}

The statistical equilibrium equations describe the radiative and collisional transitions between the different levels of a model atom \citep[see,  e.g.,][]{hubeny2014theory}. When the collisional terms dominate the rate equations, the assumption of local thermodynamical equilibrium (LTE) is usually adequate, and the atomic level population \diar{densities (hereafter population densities)} can be obtained analytically using the Saha-Boltzmann equations. When the radiative terms \diar{become relevant}, however, the radiation field greatly influences the population \diar{densities} (NLTE). Because of this cross-dependence of the population \diar{densities} with the radiation field, the NLTE problem must be solved iteratively in order to make them consistent with each other. Moreover, the nonlocality of the radiation field increases the \diar{complexity} of the problem because all grid cells must be solved simultaneously. 

Early attempts solved the rate equations using Lambda iteration, which is based on the fixed-point iteration method \citep[see, e.g.,][]{hubeny2014theory}. However, this scheme has very poor convergence properties and cannot be used in practice. \citet{1969ApJ...158..641A} proposed a complete linearization method (of the second-order transfer equation) to solve the structure and radiation emerging from static stellar atmospheres. In their implementation, physical accuracy was neglected to make the problem computationally tractable, but it inspired future developments in the field (see below). A different approach, introduced by \citet{1972lfpm.conf..145R}, used the core saturation approximation to eliminate passive photon scatterings in the line core, while only keeping the much more efficient scatterings in the line wings. \diar{The latter} improved the conditioning of the rate equations, allowing traditional Lambda iteration to converge in a reasonable (but large) number of iterations. 

The most successful methods for solving the statistical equilibrium equations are based on the operator splitting technique \citep{1973ApJ...185..621C} combined with a linearization of the problem \diar{\citep[e.g.,][]{1969ApJ...158..641A,SCHARMER198556,1992A&A...262..209R}}. \citet{SCHARMER198556} linearized the first-order radiative transfer equation and the rate equations with respect to the population \diar{densities} until they were able to derive a linear system to estimate a correction. \diar{\citep[][RH92 hereafter]{1992A&A...262..209R}} followed a slightly different approach and replaced some of the quantities that depend on the population \diar{densities} with the value from the previous iteration. The fundamental difference between these two methods is that the complete linearization method of \citet{SCHARMER198556} is a minimization method of the error in the rate equations, whereas the RH92 method is closer to the fixed-point iteration method, but uses the operator-splitting technique to drive the solution. Furthermore, the complete linearization method of \diar{\citet{SCHARMER198556}} operates on the source function, whereas the RH92 method operates on the emissivity, allowing for a simpler treatment of overlapping (active) transitions and partial redistribution effects \citep{Uitenbroek_2001,2012A&A...543A.109L,2017A&A...597A..46S}. 

The performance of these methods is largely determined by the choice of the approximate operator. The simplest block-diagonal (local) operator \citep{1986JQSRT..35..431O} decouples the explicit dependence of the rate equations with respect to space, and it requires very little storage and operations, making it a great choice for multidimensional problems \citep[e.g.,][]{2009ASPC..415...87L, 2018A&A...615A.139A}. However, information about the \diar{nonlocal} contribution to the intensity is neglected, as is its origin. A better prediction of the mean intensity can be attained by using the single-point quadrature global operator \citep{1981ApJ...249..720S,1982StoOR..19.....S}, which was greatly inspired by the Eddington-Barbier approximation \citep{1921MNRAS..81..361M,1926ics..book.....E,1943AnAp....6..113B}. Although it only requires one coefficient per ray and direction (two, when linear interpolation is used), the equations are again spatially coupled, and they must be solved together. Schemes using the global operator generally converge in fewer iterations than those using the local operator. \diar{Several codes with implementations of} the complete linearization \citep{1986UppOR..33.....C,1995ApJ...455..376H} and RH92 \citep{Uitenbroek_2001,2009ASPC..415...87L,2015A&A...574A...3P,2015A&A...577A...7S,2018A&A...615A.139A,2018A&A...617A..24M,2021ApJ...917...14O} methods have been extensively used by the solar and stellar communities. 

A \diar{common} way of solving nonlinear systems of equations is the Newton-Raphson method \citep{Raphson2021-RAPAAU-2,newton1736method}. The main limitation to applying it to the statistical equilibrium equations is the expensive calculation of the Jacobian matrix that is required in each iteration. In this paper, we propose to use a modification of the Newton-Raphson method known as the Jacobian-free Newton-Krylov method \citep{KNOLL2004357}, to solve the radiative transfer problem. In this method, the Jacobian matrices are neither inverted nor built or stored. Instead, an iterative inversion solver based on Krylov subspaces \citep{krylov1931numerical} is used to estimate the Newton\diar{-Raphson} correction to the unknowns. This method has already proved to be efficient in several fields, such as hydrodynamics or neutron scattering problems. Compared to the method of \citet{SCHARMER198556}, our method does not require any explicit linearization of the rate and radiative transfer equations, and it does not use the operator splitting.

\diar{In Sect.~\ref{sec:met} we introduce the numerical problem under consideration and the proposed numerical method for its resolution. In Sect.~\ref{sec:res} we discuss our results, and in Sect.~\ref{sec:con} we summarize our conclusions and discuss potentially interesting developments for future studies.}


\section{\diar{Problem and methods}}\label{sec:met}

\subsection{Mathematical description of the problem}
\label{part:problem}

\diar{\subsubsection{Theory}}

We adopt the notation used in \cite{Uitenbroek_2001} to express the statistical equilibrium equations. \diar{We furthermore assume a plane-parallel geometry hereafter}. In all equations, unless mentioned otherwise, lower indices refer to atomic levels and upper indices refer to depth points within the atmosphere. The RH92 notation elegantly unifies the expressions for bound-bound and bound-free transitions, allowing for a very clean implementation of the rate equations. For a bound-bound transition between a lower level $i$ and upper level $j$, we can define
\begin{eqnarray}
V_{ij} &=& \frac{h\nu}{4\pi}B_{ij}\phi_{ij}(\nu,\mu),\label{eq:Vij}\\
V_{ji} &=& \frac{h\nu}{4\pi}B_{ji}\psi_{ij}(\nu,\mu),\\
U_{ji} &=& \frac{h\nu}{4\pi}A_{ji}\psi_{ij}(\nu,\mu),
\end{eqnarray}
where $A_{ji}$, $B_{ji}$, and $B_{ij}$ are the Einstein coefficients, and $\phi_{ij}$ and $\psi_{ij}$ are the line absorption and emission profiles. \diar{$\nu$ is the frequency, and $\mu$ is the line-of-sight angle cosine.} Similarly, for a bound-free transition, we can define
\begin{eqnarray}
V_{ij} &=& \alpha_{ij}(\nu),\label{eq:Vij2}\\
V_{ji} &=& n_e \Phi_{ij}(T)\exp\left\{-\frac{h\nu}{k_BT}\right\}\alpha_{ij}(\nu),\\
U_{ji} &=& V_{ji}\left(\frac{2h\nu^3}{c^2}\right),
\end{eqnarray}
where $\alpha_{ij}$ is the photoionization cross-section, $n_e$ is the electron density, and $\Phi_{ij}(T)$ is the Saha-Boltzmann function \diar{evaluated at temperature $T$,}
\diar{
\begin{eqnarray}
\Phi_{ij}(T) &=& \frac{g_i}{2g_j}\bigg(\frac{h^2}{2\pi m_e k_BT}\bigg)^{3/2}\exp\bigg\{\frac{E_j-E_i}{k_BT}\bigg\},
\end{eqnarray}
}
\diar{where $g$ denotes the level statistical weight and $E$ the level energy, and $m_e$ is the mass of the electron}. In practice, these expressions can be further simplified using the Einstein relations between coefficients, so that we obtain
\begin{eqnarray}
V_{ji} &=& g_{ij} V_{ij},\\
U_{ji} &=& g_{ij}\left(\frac{2h\nu^3}{c^2}\right)V_{ij}.
\end{eqnarray}
For bound-bound transitions, assuming complete-redistribution of scattered photons, $g_{ij}=g_i/g_j$. For bound-free transitions,
\begin{equation}
\label{eq:gij}
g_{ij} = n_e\Phi_{ij}(T)\exp\left\{-\frac{h\nu}{k_BT}\right\} = \frac{n_i^*}{n_j^*}\exp\left\{-\frac{h\nu}{k_BT}\right\},
\end{equation}
where the asterisk-superscript denotes the LTE \diar{atomic level} population. 

We can now write the rate equations as a function of $V_{ij}$, regardless of whether we consider bound-bound or bound-free transitions. We recall the rate equation for the atomic level $i$ at depth index $k$, 
\begin{equation}
\label{eq:rate_eq}
\sum_p \bigg\{ n_p^k (C_{pi}^k+R_{pi}^k) \bigg\} = \sum_p \bigg\{ n_i^k (C_{ip}^k+R_{ip}^k)\bigg\},
\end{equation}
where $n_i^k$ is the population density of the atomic level $i$ at depth index $k$. $C_{ij}^k$ and $R_{ij}^k$ are the collisional and radiative rate coefficients of the transition $i \rightarrow j$ at depth index $k$ with $C_{ii}^k = R_{ii}^k = 0$. The radiative rate coefficients can be expressed as a double integral over the angle and frequency of the intensity \citep{Uitenbroek_2001},
\begin{eqnarray}
\label{eq:Rij}
R_{ij}^k &=& \frac{1}{2}\int_{-1}^{1} d\mu \int_{0}^{\infty} \frac{d\nu}{h\nu}V_{ij}^k I^k_{\mu\nu}(\vec n) \quad i < j \\
\label{eq:Rji}
R_{ji}^k &=& \frac{1}{2}\int_{-1}^{1} d\mu \int_{0}^{\infty} \frac{d\nu}{h\nu}\bigg[\bigg(\frac{2h\nu^3}{c^2}\bigg)+I^k_{\mu\nu}(\vec n)\bigg] g_{ij}^k V_{ij}^k \quad i < j,
\end{eqnarray}
for a plane-parallel atmosphere. The expressions for $V_{ij}^k$ and $g_{ij}^k$ are given in Eqs.~\ref{eq:Vij}-\ref{eq:gij} for bound-bound and bound-free transitions. The last component $I_{\mu\nu}^k$ is the intensity in the direction $\mu$ at a frequency $\nu$ and at the depth index $k$. \diar{The vector $\vec n$ contains the population densities with the chosen structure $(n_1^1,...,n_{N_\ell}^1,~...~,n_1^{N_z},...,n_{N_\ell}^{N_z})^\textrm{T}$, where $N_z$ and $N_\ell$ are the number of depth points and active atomic levels, respectively.} While all the other quantities do not depend on the population densities, the intensity involves them all in a nonlinear and nonlocal fashion through the radiative transfer equation (RTE),
\begin{eqnarray}
\label{eq:RTE1}
I_{\mu\nu}(\tau_{\mu\nu}) &= \int_{\tau_{\mu\nu}}^{\infty}S_{\mu\nu}(t)e^{-(t-\tau_{\mu\nu})}dt \quad& \mu > 0 \\
\label{eq:RTE2}
I_{\mu\nu}(\tau_{\mu\nu}) &= \int_{\tau_{\mu\nu}}^0 S_{\mu\nu}(t)e^{\diar{-(\tau_{\mu\nu}-t)}}dt \quad& \mu < 0,
\end{eqnarray}
where $S_{\mu\nu} = \eta_{\mu\nu}/\chi_{\mu\nu}$ is the source function, $\chi_{\mu\nu}$ and $\eta_{\mu\nu}$ are the total opacity and emissivity, respectively, which can be calculated through
\begin{eqnarray}
\label{eq:chi}
\chi_{\mu\nu} &=& \chi_c + \chi_\textrm{sca} + \sum_i\sum_{j>i}V_{ij}\bigg(n_i-g_{ij}n_j\bigg) \\
\label{eq:eta}
\eta_{\mu\nu} &=& \eta_c + \chi_\textrm{sca}J_\nu+\sum_j\sum_{i<j}\bigg(\frac{2h\nu^3}{c^2}\bigg)g_{ij}V_{ij}n_j,
\end{eqnarray}
where the subscript $c$ refers to the background continuum contribution, and sca indicates the background-scattering contribution, which are assumed to be independent with respect to the active population densities. \diarp{The mean intensity $J_\nu$ can be computed using}
\begin{eqnarray}
J_\nu &=& \frac{1}{2}\int_{-1}^{1}I_{\mu\nu}(\vec n)d\mu.
\end{eqnarray}
The presence of $J_\nu$ in the scattering term of Eq.~\ref{eq:eta} might cause the calculations to become more complex because it depends on the intensity, which in turn depends on opacities and emissivities. Because these scattering terms do not originate from active transitions of the atom, however, we use a previous estimation of the mean intensity $J_\nu^\dagger$ instead of $J_\nu$ in Eq.~\ref{eq:eta}.
The optical thickness $\tau_{\mu\nu}$ is obtained by integrating the opacity over depth,
\begin{equation}
\tau_{\mu\nu}(z) = \frac{1}{|\mu|}\int_{0}^{z}\chi_{\mu\nu}(z')dz' > 0.
\end{equation}

Equation \ref{eq:rate_eq} describes a system of \diar{$N_\ell\times N_z$} equations and variables $n_i^k$ in which $N_z$ equations are redundant. Therefore, we replace one rate equation by a particle conservation equation per depth-point,
\begin{equation}
\sum_{p}n_p^k = n_\textrm{tot}^k,
\end{equation}
where $n_\textrm{tot}^k$ is the total atom density at the depth index $k$ and is kept constant. The replacement was made on the most populated atomic level at each depth point for numerical stability purposes. 

Finally, the system of equations was reformulated as
\begin{equation}
\label{eq:residuals_1}
F_i^k(\vec n) \stackrel{\text{def}}{=} \sum_p \bigg\{ n_i^k (C_{ip}^k+R_{ip}^k) - n_p^k (C_{pi}^k+R_{pi}^k)\bigg\}
\end{equation}
for radiative rate equations, and
\begin{equation}
\label{eq:residuals_2}
F_i^k(\vec n) \stackrel{\text{def}}{=} n_\textrm{tot}^k - \sum_{p}n_p^k
\end{equation}
for mass conservation equations. Altogether, \diarp{Eqs.~\ref{eq:residuals_1}-\ref{eq:residuals_2}} form a residual vector $\vec F(\vec n)$ with the same \diar{dimension} as $\vec n$. Solving the system of equations for the vector of \diar{population} densities $\vec n$ is therefore equivalent to finding the root of the \diar{residual} vector $\vec F$. The calculation of $\vec F$ for a given atmosphere and population densities is detailed in algorithm \ref{alg:vec_F}. The residual \diar{vector} is the central part of the solving process and constitutes the main computational cost of it. Thus, we evaluated the performance of a solver by the number of computation calculations of $\vec F$ (hereafter calls) needed to solve the problem to a given precision.

\begin{algorithm}
\caption{Calculation of the residual \diar{vector} $\vec F$}\label{alg:vec_F}
\KwData{a population \diar{density} vector $\vec n$, an estimation of $\vec J$}
\KwResult{$\vec F(\vec n)$, possibly a new estimate of $\vec J$}

\For{$\nu = \nu_1,...,\nu_{N_\nu}$}
{
    $\vec J_\nu^\dagger \gets \vec J_\nu$\;
    \If{updating $J$}{$\vec J_\nu \gets \vec 0$\;}

    \For{$\mu = \mu_1,...,\mu_{N_\mu}$}
    {
        $\vec \chi_{\mu\nu} \gets $ Eq. \ref{eq:chi}\; $\vec \eta_{\mu\nu} \gets $ Eq. \ref{eq:eta}\;
        $\vec S_{\mu\nu} \gets \vec \eta_{\mu\nu}/\vec \chi_{\mu\nu}$\;
        $\vec I_{\mu\nu} \gets $ Eq. \ref{eq:Iplus} and \ref{eq:Iminus}\;
        \For{transitions $i\rightarrow j$ with $j < i$\diar{, for all $z$}}
        {
            $\vec R_{ij} \gets \vec R_{ij} + (\mu,\nu)$-contribution of Eq. \ref{eq:Rij}\;
            $\vec R_{ji} \gets \vec R_{ji} + (\mu,\nu)$-contribution of Eq. \ref{eq:Rji}\;
        }
        \If{updating $J$}{$\vec J_\nu \gets \vec J_\nu + \omega_{\mu}\vec I_{\mu\nu}$\;}
        
    }
}
$\vec F \gets \vec 0$\;
\For{transitions $i\rightarrow j$ with $j < i$}
{
    $\vec \kappa \gets \vec n_i (\vec C_{ij}+\vec R_{ij})-\vec n_j(\vec C_{ji}+\vec R_{ji})$\;
    $\vec F_i \gets \vec F_i + \vec \kappa$\;
    $\vec F_j \gets \vec F_j - \vec \kappa$\;
}
\For{$k = 1,...,N_z$}
{
    $i \gets \{j~;~n_j^k = \max_p(n_p^k)\}$\;
    $F_i^k \gets n_\textrm{tot}^k-\sum_p n_p^k$\;
}
\end{algorithm}

\diar{\subsubsection{Discretization of the RTE}}

\diar{When the radiative rates are computed, the angular and frequency integrals are in practice discretized according to quadrature schemes and yield quadrature coefficients $(\omega_\mu,\omega_\nu)$ for each set $(\mu,\nu)$.
Equations \ref{eq:RTE1} and \ref{eq:RTE2} are discretized along the depth axis, and the involved integrals can be calculated assuming a depth-dependent profile for $S_{\mu\nu}$. This profile is usually taken as simple piece-wise polynomial functions. We considered piece-wise linear functions \citep{1987JQSRT..38..325O}, which yielded
\begin{eqnarray}
\label{eq:Iplus}
I_{\mu\nu}^k =& I_{\mu\nu}^{k+1}e^{-\Delta\tau_{\mu\nu}^k}+a^kS_{\mu\nu}^k+b^kS_{\mu\nu}^{k+1} \quad& \mu > 0 \\
\label{eq:Iminus}
I_{\mu\nu}^k =& I_{\mu\nu}^{k-1}e^{-\Delta\tau_{\mu\nu}^{k-1}}+a^{k-1}S_{\mu\nu}^k+b^{k-1}S_{\mu\nu}^{k-1} \quad& \mu < 0,
\end{eqnarray}
where the coefficients $a^k$ and $b^k$ are given in Appendix \ref{ap:piece}, and
\begin{equation}
\Delta\tau_{\mu\nu}^k \approx \frac{1}{2\diar{|\mu|}}|z^k-z^{k+1}|(\chi_{\mu\nu}^k+\chi_{\mu\nu}^{k+1})
\end{equation}
is the optical thickness of the slab.} 

\diar{At the top of the atmosphere, we assumed that there is no incoming radiation. The deepest point in the atmosphere was assumed to be thermalized so that the intensity at this location is the Planck function $B_\nu$ \diarp{at the local temperature $T$}},
\begin{eqnarray}
I_{\mu\nu}^{N_z} =& B_\nu(T^{N_z}) \quad& \mu > 0 \\
I_{\mu\nu}^1 =& 0 \quad& \mu < 0.
\end{eqnarray}

\diar{The angular integrals were evaluated using a Gauss-Legendre quadrature defined in $]0,1[$. \diarp{At each depth index $k$, the incoming and outgoing rays were considered in the calculation of the mean intensity $\vec J$}. The number of quadrature points is a parameter set by the user. We normally ran with five quadrature points with two rays per angle.}
\\

\subsection{Newton-Raphson method}

\subsubsection{Basics}

\begin{figure}[!ht]
    \centering
    \includegraphics[scale=0.44]{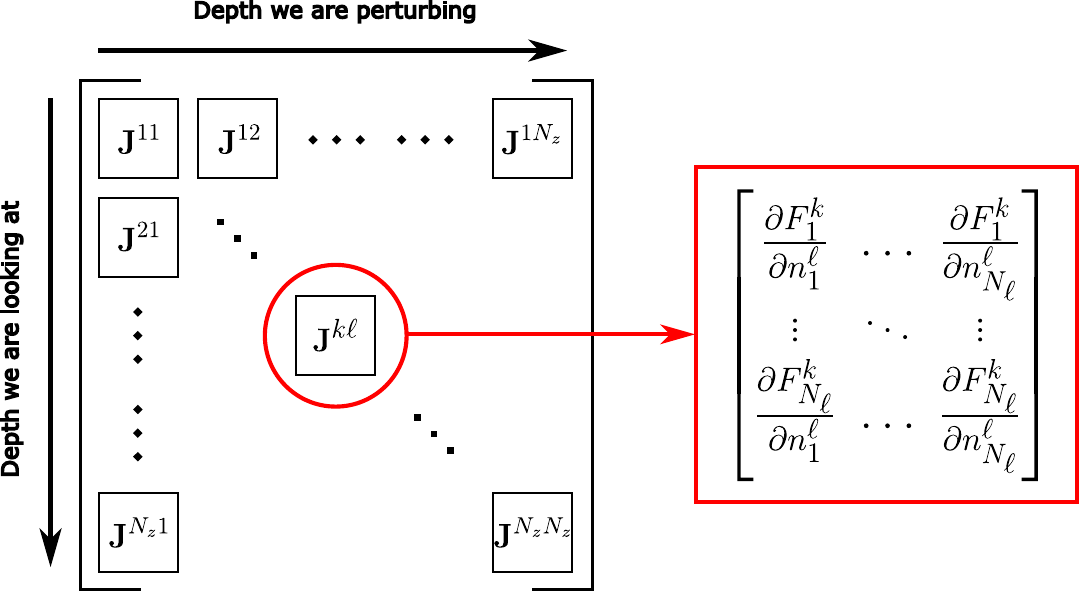}
    \caption{Jacobian matrix structure for a \diar{$N_\ell$-level} atom problem. $\bold{J}_F$ is a \diar{$(N_\ell N_z)\times(N_\ell N_z)$} matrix that contains the derivatives of the residual vector components with respect to the population densities. This matrix can be considered as a $N_z\times N_z$ block matrix, where each block \diar{$\bold{J}^{k\ell}$ is a $N_\ell\times N_\ell$} matrix. \diar{$\bold{J}^{k\ell}$} stores the derivatives of $\vec F$ at depth index $k$ with respect to the population densities at \diar{depth index $\ell$}.}
    \label{fig:jac_example}
\end{figure}

The system of nonlinear equations $\vec F(\vec n) = \vec 0$ for the vector $\vec n$ may be solved through several numerical iterative methods. The Newton\diar{-Raphson} method is one of the simplest and most powerful ones \diar{\citep[e.g.,][]{2002nrca.book.....P}}. When $\vec n^{(p)}$ is the estimate of the solution on the $p^\textrm{th}$ iteration, the next iterate $\vec n^{(p+1)}$ is sought such that $\vec F(\vec n^{(p+1)}) = \vec 0$. When we further define $\delta \vec n^{(p)} = \vec n^{(p+1)} - \vec n^{(p)}$ as the $p^\textrm{th}$ incremental, the Newton\diar{-Raphson} method relies on a linearization of $\vec F(\vec n^{(p+1)})$,

\begin{equation}
\vec 0 = \vec F(\vec n^{(p+1)}) = \vec F(\vec n^{(p)} + \delta\vec n^{(p)}) \approx \vec F(\vec n^{(p)})+\bold{J}_F(\vec n^{(p)})\delta\vec n^{(p)},
\end{equation}
where we have introduced the Jacobian matrix $\bold{J}_F$ associated with the residual \diar{vector} $\vec F$ and evaluated at $\vec n^{(p)}$. \diar{A possible representation of $\bold{J}_F$} is given in Fig.~\ref{fig:jac_example}.
Solving the latter linear system for $\delta \vec n^{(p)}$ yields
\begin{eqnarray}
\label{eq:newton_approx}
\delta\vec n^{(p)} = -\bold{J}_F^{-1}(\vec n^{(p)})\vec F(\vec n^{(p)}),
\end{eqnarray}
from which we may compute the next iterate $\vec n^{(p+1)}$. This new estimate is a priori not a solution to $\vec F(\vec n) = \vec 0$, although the iterative process ensures a quadratic convergence to a solution in the best cases \diar{\citep[e.g.,][]{doi:10.1137/1.9781611971200}}. The initial guess $\vec n^{(0)}$ might be given, for instance, by the LTE  or the \diar{radiation-free predictions of the population densities}. The linearization introduced by the Newton-Raphson method only consists of a mean to solve the raw statistical equations, whereas RH92 solves a linearized approximated version of the problem.

\subsubsection{Limitations of \diar{the Newton-Raphson} method}

We note that $\delta\vec n^{(p)}$ might also lead to a poorer estimation of the solution. This behavior can occur when the correction $\delta\vec n^{(p)}$ lies beyond the domain of linearity of the residual \diar{vector} around the evaluation vector $\vec n^{(p)}$. A simple way to overcome this behavior is to limit the incremental vector with a dampening factor $\alpha$ and try $\alpha = 1,0.5,0.25,...$ until $||\vec F(\vec n^{(p)}+\alpha \delta\vec n^{(p)})|| < ||\vec F(\vec n^{(p)})||$. This procedure is the simplest of the so-called line-search methods, although more elaborated ones exist \citep[see][for instance]{doi:10.1137/1.9781611971200}. 

A second problem deals with the possibility to produce solution estimates with negative entries. While mathematically correct, a solution estimate \diar{with negative population densities} is physically incorrect, and the solver may even overflow when solving the RTE. A possible solution to prevent negative entries consists of limiting the correction at each depth independently. 

A third and inherent problem of the Newton\diar{-Raphson} method deals with the quality of the initial guess. The initial guess is usually an important factor in the method convergence \diar{rate} or even the failure of the method. The method does not converge or ventures outside the domain of definition of $\vec F$ when a poor starting point is given. The method may also be trapped in a local minimum of the residual \diar{vector}, which can be difficult to spot. Several tools such as continuation methods can be used to build a robust solver based on the Newton\diar{-Raphson} method. More details are given in \cite{KNOLL2004357}.


\diar{The Newton-Raphson} requires the knowledge of \diar{the inverse of a Jacobian matrix $\bold{J}_F^{-1}$ at each iterative step}. Several issues can potentially arise when these quantities are computed, and we list them below.
\begin{enumerate}
    \item Implementation: Most problems do not have an analytical expression for $\bold{J}_F$, and an approximation needs to be given (e.g., by finite differences). The convergence is therefore likely to be less than quadratic. In the worst case, the method may fail when the approximation is too coarse. 
    \item Storage: For large problems, storing $\bold{J}_F$ can be problematic. 
    \item Time consumption: Inversion of $\bold{J}_F$ quickly becomes time-consuming as the size of the problem increases, considering traditional inversion routines such as Gauss-Jordan elimination. The computation of the full Jacobian might also be expensive.
\end{enumerate}
The radiative transfer problem we considered disqualifies a classical Newton\diar{-Raphson} method mainly because of the computational cost derived from building of $\bold{J}_F$, even when using analytical expressions. The Newton\diarp{-Raphson} method therefore requires the information of the Jacobian matrices without building them or only building them partially, in order to keep an efficient solver. The next sections detail a way to bypass this thorny problem.

\subsection{Iterative inversion: Krylov methods}
\label{part:Krylov}

Equation \ref{eq:newton_approx} \diar{is equivalent to} a linear system of the form $\bold{A}\vec x = \vec b$, where $\bold{A} = \bold{J}_F(\vec n^{(p)})$, $\vec x = \delta\vec n^{(p)}$ and $\vec b = -\vec F(\vec n^{(p)})$. This linear system can be solved for $\vec x$ without inverting $\bold{A}$ using iterative approaches such as Krylov methods. In short, Krylov methods are used to solve large linear systems through projections onto a Krylov subspace $K_s$,
\begin{equation*}
K_s = \textrm{span}(\vec r_0, \bold{A}\vec r_0, \bold{A}^2\vec r_0, ..., \bold{A}^{s-1}\vec r_0),
\end{equation*}
where $\vec r_0 = \vec b - \bold{A}\vec x_0$ is the initial residual vector built from the initial guess $\vec x_0$. Since $\vec x$ is meant to represent a Newton\diar{-Raphson} correction, a typical initial guess would be zero \diar{and thus $\vec r_0 = \vec b$} (we considered that no correction is required initially). Another initial condition might be given by the solution to $\bold{P}\vec x_0 = \vec b$, where $\bold{P}$ is a preconditioner \diar{(Sect.~\ref{part:prec1})}. Then, the solution is estimated as
\begin{equation}
\vec x - \vec x_0 = \sum_{i=1}^{s}\kappa_i \vec q_i,
\end{equation}
where the set of vectors $(\vec q_1, ..., \vec q_s)$ is a basis of $K_s$ and $(\kappa_1, ..., \kappa_s)$ are the corresponding coordinates. The purpose of a Krylov method is therefore to construct a basis of $K_s$, and then to determine the corresponding scalars $\kappa_i$ through projection methods. This construction is made iteratively in $s$ iterations (one iteration per basis vector). Each iteration adds a new component to the solution estimate. The solution is sought such that $||\vec r||_2 = ||\vec b-\bold{A}\vec x||_2 < \delta ||\vec r_0||_2$, where $\delta$ is a relative tolerance set by the user. We note that the given tolerance might be achieved in fewer than $s$ iterations. 

The size of the Krylov subspace is related to the precision of the solution that can be achieved, the latter also depending on the method employed. If $s$ is too small, the desired tolerance level might not be achieved. On the other hand, when $s$ is chosen to be equal to the size of $A$, a Krylov method will eventually converge to the exact solution in theory. In practice, round-off and truncation errors will limit the maximum precision that can be expected. From the definition of $K_s$, we note that only matrix-vector products are required in these methods, which is the keystone of Sect.~\ref{part:JFNK}. 

A plethora of Krylov methods has been developed over the past decades, among which two popular techniques and their respective variants are broadly used in various physics problems. We briefly describe them below.
\begin{enumerate}
    \item The generalized minimal residual method (GMRES) is usually based on the Arnoldi process or Householder transforms to produce orthonormalized bases. It was first developed in \cite{doi:10.1137/0907058} as an improvement and an extension to nonsymmetric matrices of the MINRES method developed by \cite{doi:10.1137/0712047}. It is a very versatile linear system solver.
    \item The bi-conjugate gradient stabilized (BiCGSTAB) is based on the Lanczos bi-orthogonalization procedure, which generates nonorthogonal bases. It was first developed in \cite{doi:10.1137/0913035} as a variant of the biconjugate gradient method (BiCG).
\end{enumerate}
Both methods can be used with any invertible matrix. GMRES requires one matrix-vector product per iteration, whereas two are needed with BiCGSTAB, but the latter requires less memory than GMRES. This is especially true whenever the number of iterations required for convergence is large because BiCGSTAB uses a constant amount of memory per iteration and GMRES does not. The most crucial feature of GMRES is a strict decrease of the residual norm $||\vec r||_2$ throughout the iterative process, whereas the convergence behavior of BiCGSTAB is less regular \citep{doi:10.1137/0913035}. 

\subsection{Jacobian-free Newton-Krylov methods}
\label{part:JFNK}

\subsubsection{Setup}

We still did not address the problem of the Jacobian matrix estimation and the potential high computational cost it represents. Fortunately, Krylov methods applied to the Newton\diar{-Raphson} iteration (Eq.~\ref{eq:newton_approx}) only require the action of the Jacobian matrix $\bold{J}_F$ onto a \diar{generic} vector $\vec v$ (see Sect.~\ref{part:Krylov}). The operation $\bold{J}_F(\vec n^{(p)})\vec v$ can be approximated using finite differences \diar{\citep[e.g.,][]{KNOLL2004357}},
\begin{eqnarray}
\bold{J}_F(\vec n^{(p)})\vec v =& \frac{\vec F(\vec n^{(p)} + \epsilon \vec v)-\vec F(\vec n^{(p)})}{\epsilon} + \mathcal{O}(\epsilon) \quad& \textrm{forward} \label{eq:fordif}\\
\bold{J}_F(\vec n^{(p)})\vec v =& \frac{\vec F(\vec n^{(p)})-\vec F(\vec n^{(p)} - \epsilon \vec v)}{\epsilon} + \mathcal{O}(\epsilon) \quad& \textrm{backward} \\
\bold{J}_F(\vec n^{(p)})\vec v =& \frac{\vec F(\vec n^{(p)} + \epsilon \vec v)-\vec F(\vec n^{(p)} - \epsilon \vec v)}{2\epsilon} + \mathcal{O}(\epsilon^2) \quad& \textrm{central} \label{eq:centdif},
\end{eqnarray}
where $\epsilon$ is the difference step. These schemes do not use the Jacobian matrix, but rather extra calls of $\vec F$, which is a huge computational and storage gain especially for large problems, but at the cost of precision. This is the keystone of Jacobian-free Newton-Krylov solvers (hereafter JFNK). These methods were first presented by \cite{KNOLL2004357}. Since the \diar{residual} vector $\vec F(\vec n^{(p)})$ is already computed and passed to the Krylov solver, first-order schemes \diar{(forward and backward)} only require one fresh call of $\vec F$. In comparison, the second-order scheme \diar{(central)} requires two fresh calls of $\vec F$, which is a major drawback. In our problem, every evaluation of $\vec F$ translates into solving the RTE for all rays at all frequencies for a given $\vec n$.
We also note that the finite-differences calculations in Eq.~\ref{eq:fordif}-\ref{eq:centdif} do not estimate the individual elements of the Jacobian matrix, but are instead used to directly estimate the matrix-vector product. 

Since the Newton\diar{-Raphson} method is iterative, the matrix-vector product estimates are only needed to be accurate enough to guarantee convergence. This is the main reason why the vast majority of JFNK solvers use first-order schemes rather than higher-order ones \citep{KNOLL2004357}. 
In practice, round-off and truncation errors may occur, and an optimal choice of $\epsilon$ is hard to find. The first source of error is caused by the finite arithmetic precision of computers, and the second source of error is due to the limited accuracy of the scheme. Several empirical choices for $\epsilon$ are further detailed in \cite{KNOLL2004357} to minimize both sources of error. 

\subsubsection{Augmentation of numerical precision with complex numbers}
\begin{figure*}[!ht]
\centering
\includegraphics[width=0.99\textwidth]{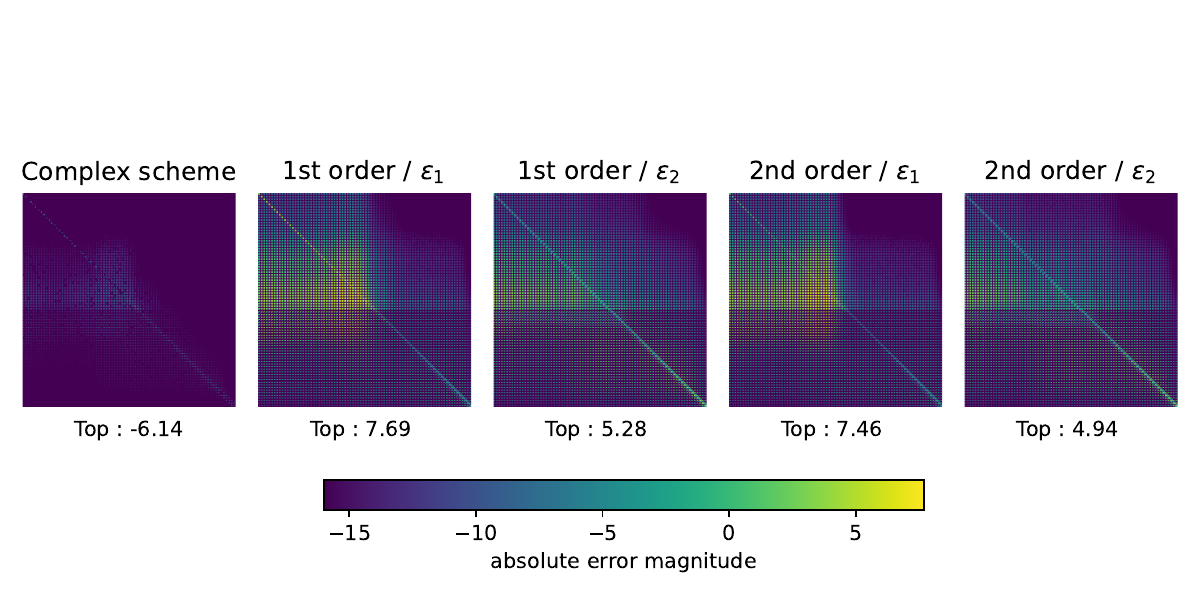}
    \diar{\caption{Jacobian matrix estimation error for the three-level \ion{Ca}{ii} setup, evaluated at the LTE populations and for the different schemes. \diarp{The first-order scheme is the forward one.} The difference steps $\epsilon_1$ and $\epsilon_2$ are common choices \citep{KNOLL2004357} and read $\epsilon_1 = \sum_{i,k=1}^{N_\ell,N_z}b|(n_i^k)^{(p)}|/(N_\ell N_z||\vec v||_2)+b$, where $b = 10^{-6}$, and $\epsilon_2 = \epsilon_\textrm{mach}^{1/r}\sqrt{1+||\vec n^{(p)}||_2}/||\vec v||_2$, where $\epsilon_\textrm{mach}$ is the machine epsilon for double-precision numbers, $r = 2$ for forward and backward schemes, and $r = 3$ for the central scheme. The logarithm of the absolute error (color bar) is clipped below $-16$.} 
\label{fig:scheme_perf}}
\end{figure*}

For the considered problem, the components of $\vec n^{(p)}$ cover a wide range of values\diar{. For instance,} considering a two-level hydrogen atom in a FAL-C atmosphere \citep{1993ApJ...406..319F}, the LTE population densities cover the range $10^{4}$ up to $10^{17}$ cm$^{-3}$. This leads to high values of $\epsilon$ considering the empirical choices. A rescaling of the densities is possible to keep reasonable $\epsilon$ values. However, the large span of densities remains problematic within the residual \diar{vector} $\vec F$ itself and leads to round-off errors. Thus, the usual numerical derivative schemes are not suited for the given problem. Fortunately, there is an alternative scheme that uses complex numbers to increase the precision and dynamic range of the calculations, which is far less affected by round-off errors or substractive cancellations \citep[e.g.,][]{KAN2022113732,Martins_Ning_2021},
\begin{equation}
\label{eq:complex}
\bold{J}_F(\vec n^{(p)})\vec v = \frac{\mathfrak{Im}[\vec F(\vec n^{(p)} + i\epsilon \vec v)]}{\epsilon} + \mathcal{O}(\epsilon^2),
\end{equation}
where $i$ is the imaginary unit. This special scheme only requires one fresh call of $\vec F$ and is \diar{second-order} accurate as the central difference scheme. It requires setting up the main routines for complex arithmetic operations, however. 
We used the linear piece-wise source function scheme described in Sect.~\ref{part:problem} to integrate the RTE along rays, for which the modifications are straightforward. The remaining truncation error can be greatly attenuated by selecting a tiny $\epsilon$ value. \cite{Martins_Ning_2021} recommend a typical value $\epsilon \sim 10^{-200}$ for double-precision functions. The imaginary part is only used for the Krylov solver, otherwise only the (unperturbed) real part is considered. \diar{Since the imaginary part is typically very small compared to the real part, equations that involve the perturbations should be linearized (e.g., by computing the source function). This prevents introducing undesired arithmetic errors. Figure~\ref{fig:scheme_perf} points out a typical accuracy issue in the computation of the Jacobian matrices when using traditional schemes. The complex scheme alone can provide accurate estimates of Jacobian matrices for our given NLTE problem. For this sole reason, we used the complex scheme in our JFNK solvers.}

\subsubsection{\diar{Preconditioning}}
\label{part:prec1}
Although we have introduced an accurate method to evaluate the matrix-vector products $\bold{J}_F(\vec n^{(p)})\vec v$, the Jacobian matrix may potentially be ill-conditioned. Consequently, the Krylov solver may require many iterations to converge to the desired tolerance because of the high condition number of $\bold{J}_F(\vec n^{(p)})$. Therefore, we propose preconditioning the Krylov solver in order to increase its efficiency. The preconditioning process consists of using a preconditioning matrix $\bold{P}$ (the preconditioner) such that $\bold{J}_F(\vec n^{(p)})\bold{P}^{-1}$ (right \diar{preconditioning}) or $\bold{P}^{-1}\bold{J}_F(\vec n^{(p)})$ (left \diar{preconditioning}) has a lower condition number than $\bold{J}_F(\vec n^{(p)})$. The system to be solved by the Krylov solver is no longer given by Eq.~\ref{eq:newton_approx} but rather by
\begin{equation}
\label{eq:prec}
(\bold{J}_F(\vec n^{(p)})\bold{P}^{-1})(\bold{P}\delta \vec n^{(p)}) = -\vec F(\vec n^{(p)})
\end{equation}
for right preconditioning and by
\begin{equation}
\label{eq:prec2}
(\bold{P}^{-1}\bold{J}_F(\vec n^{(p)}))\delta \vec n^{(p)} = -\bold{P}^{-1}\vec F(\vec n^{(p)})
\end{equation}
for left preconditioning. The preconditioned system is expected to be solved in fewer iterations than the original system. Equation \ref{eq:prec} is solved for $\vec y = \bold{P}\delta \vec n^{(p)}$ first then for $\delta \vec n^{(p)}$. \diarp{If the right preconditioning is used, the matrix-vector product scheme (Eq.~\ref{eq:complex}) can be adapted for Eq.~\ref{eq:prec} and reads}
\begin{equation}
\bold{J}_F(\vec n^{(p)})(\bold{P}^{-1}\vec v) = \frac{\mathfrak{Im}[\vec F(\vec n^{(p)} + i\epsilon (\bold{P}^{-1}\vec v))]}{\epsilon} + \mathcal{O}(\epsilon^2).
\end{equation}
\diarpp{If the left preconditioning is chosen, Eq.~\ref{eq:complex} is used directly, and then the result is left-multiplied by $\bold{P}^{-1}$. The second member passed to the Krylov solver in this case is $-\bold{P}^{-1}\vec F(\vec n^{(p)})$.
In our JFNK solvers, we used the left preconditioning because the right preconditioning involves an additional inversion step.}
Because the preconditioner is meant to improve the overall performance of the inversion process, $\bold{P}$ should be easy to calculate and invert. A commonly used \diar{algebraic} preconditioner is given by the diagonal or block diagonal of the matrix that is to be inverted (easy inversion and matrix-vector multiplication). This simple matrix is also known as a Jacobi \diar{(or block-Jacobi)} preconditioner. This choice is particularly interesting in our case because the block diagonal of $\bold{J}_F(\vec n^{(p)})$ is as costly to compute as a call of $\vec F$ (see Sect.~\ref{part:problem} and Sect.~\ref{part:prec_JFNK}). \diar{We note, however, that other physics-based preconditioners could be very relevant for our case, as presented in \citet{2024A&A...682A..68J} for solving linear problems}. The calculation of such a preconditioner can be performed when calculating $\vec F(\vec n^{(p)})$, thus reusing most of the variables that were already computed to obtain the residual \diar{vector}. Algorithm \ref{alg:JFNK} illustrates the final JFNK solving routine.

\begin{algorithm}
\caption{JFNK solver}\label{alg:JFNK}
\KwData{an initial population \diar{densities} vector $\vec n^{(0)}$, an initial estimate of $\vec J$, a Newton\diar{-Raphson} tolerance $ftol$, a Krylov \diarp{relative} tolerance $r_{\mathrm{tol}}$, a maximum number of iterations $maxiter$}
\KwResult{A solution vector $\vec n$, a mean intensity estimation~$\vec J$}

$niter \gets 0$\;
$err \gets 1+ftol$\;
$\vec f \gets \vec F(\vec n^{(0)})$\Comment{update the preconditioner $\bold{P}$}\; 

\While{$niter < maxiter$ \textrm{and} $err > ftol$}
{
    $\vec x_0 \gets \vec 0$ or $\vec x_0 \gets -\bold{P}^{-1}\vec f$ \Comment{initial guess}\;
    $\delta \vec n \gets$ solution of Eq. \ref{eq:prec2} with a Krylov solver using Eq. \ref{eq:complex}\;
    $\vec n_{prev} \gets \vec n$\;
    $\vec n \gets \vec n+\delta\vec n$\;
    $\vec f \gets \vec F(\vec n)$\Comment{update $\vec J$ and $\bold{P}$}\;
    $err \gets ||(\vec n - \vec n_{prev})/(\vec n + \vec n_{prev})||_\infty$\;
    $niter \gets niter + 1$\;
}
\end{algorithm}

\subsection{Analytical Jacobian matrix}
\label{part:analytical_jac}

\subsubsection{Derivation}

In this part, we derive the expressions of the Jacobian matrix elements as a function of the population densities. In principle, these equations could be used to compute the fully analytical Jacobian matrix. While the calculation of the full Jacobian is expensive, it is at least important to detail these derivations for preconditioning purposes (Sect.~\ref{part:JFNK}). A more general derivation is provided in \cite{refId0} for derivatives according to any atmospheric parameter. The Jacobian element \diar{$J_{ij}^{k\ell}$} can be calculated as follows:
\diar{\begin{equation}
\label{eq:der_F}
J_{ij}^{k\ell} = \bigg(\frac{\partial F_i^k}{\partial n_j^\ell}\bigg) = \frac{\partial}{\partial n_j^\ell}\Bigg(\sum_p \bigg\{ n_i^k (C_{ip}^k+R_{ip}^k) - n_p^k (C_{pi}^k+R_{pi}^k)\bigg\}\Bigg)
\end{equation}}
unless $F_i^k$ is a particle conservation equation, in which case the Jacobian element is simply given by
\diar{\begin{equation}
J_{ij}^{k\ell} = \frac{\partial}{\partial n_j^\ell}\Bigg(n_\textrm{tot}^k-\sum_p n_p^k\Bigg) = -\delta_{k\ell},
\end{equation}}
where we used the fact that the population \diar{densities} $n_i^k$ are considered to be independent variables \diar{and that $n_\textrm{tot}^k$ is kept constant}, and therefore,
\diar{\begin{equation}
\frac{\partial n_i^k}{\partial n_j^\ell} = \delta_{k\ell}\delta_{ij}.
\end{equation}}
In Eq.~\ref{eq:der_F}, the collisional coefficients do not depend on the population densities. Applying the chain rule to Eq.~\ref{eq:der_F} yields
\diar{\begin{equation}
J_{ij}^{k\ell} = \bigg(\frac{\partial F_i^k}{\partial n_j^\ell}\bigg) = \delta_{k\ell}\Gamma_{ij}^k + A_{ij}^{k\ell}
\end{equation}}
where
\diar{\begin{eqnarray}
\Gamma_{ij}^k &=& \delta_{ij}\sum_p (C_{ip}^k+R_{ip}^k)-(C_{ji}^k+R_{ji}^k) \\
A_{ij}^{k\ell} &=& \frac{1}{2}\int_{-1}^{1} d\mu \int_{0}^{\infty} \frac{d\nu}{h\nu} \alpha_i^k \bigg(\frac{\partial I_{\mu\nu}^k}{\partial n_j^\ell}\bigg) \\
\alpha_i^k &=& \sum_{p>i}\bigg\{V_{ip}^k(n_i^k-g_{ip}^k n_p^k)\bigg\} - \sum_{p<i}\bigg\{ V_{pi}^k(n_p^k-g_{pi}^k n_i^k) \bigg\}
\end{eqnarray}}
by noting that the derivative of the radiative rates only involves the derivative of the intensity $I_{\mu\nu}^k$ with respect to \diar{$n_j^\ell$}. The extinction profile within each coefficient $V_{ip}^k$ is considered to be independent with respect to the population densities. Then, we can expand the derivative and further write
\diar{\begin{equation}
\label{eq:partial_dev}
\bigg(\frac{\partial I_{\mu\nu}^k}{\partial n_j^\ell}\bigg) = \sum_p \Bigg\{\bigg(\frac{\partial I_{\mu\nu}^k}{\partial \chi_{\mu\nu}^p}\bigg)\bigg(\frac{\partial \chi_{\mu\nu}^p}{\partial n_j^\ell}\bigg)+\bigg(\frac{\partial I_{\mu\nu}^k}{\partial \eta_{\mu\nu}^p}\bigg)\bigg(\frac{\partial \eta_{\mu\nu}^p}{\partial n_j^\ell}\bigg)\Bigg\}
\end{equation}}
because the intensity, through the RTE, is only a function of the optical depth and the source function. We further develop Eq. \ref{eq:partial_dev} and write
\diar{\begin{eqnarray}
\bigg(\frac{\partial \chi_{\mu\nu}^p}{\partial n_j^\ell}\bigg) &=& \delta_{p\ell}\beta_j^\ell \\
\label{eq:der_eta_n}
\bigg(\frac{\partial \eta_{\mu\nu}^p}{\partial n_j^\ell}\bigg) &=& \delta_{p\ell}\gamma_j^\ell + \chi_\textrm{sca}^p\bigg(\frac{\partial J_\nu^p}{\partial n_j^\ell}\bigg)
\end{eqnarray}}
where
\diar{\begin{eqnarray}
\beta_j^\ell &=& \sum_{s>j}V_{js}^\ell(1-\ g_{js}^\ell) \\
\gamma_j^\ell &=& \sum_{s<j}\bigg(\frac{2h\nu^3}{c^2}\bigg)V_{sj}^\ell g_{sj}^\ell,
\end{eqnarray}}
so that  Eq.~\ref{eq:partial_dev} therefore reduces to
\diar{\begin{equation}
\label{eq:partial_dev_2}
\bigg(\frac{\partial I_{\mu\nu}^k}{\partial n_j^\ell}\bigg) = \beta_j^\ell\bigg(\frac{\partial I_{\mu\nu}^k}{\partial \chi_{\mu\nu}^\ell}\bigg)+\gamma_j^\ell\bigg(\frac{\partial I_{\mu\nu}^k}{\partial \eta_{\mu\nu}^\ell}\bigg)+\sum_p \chi_\textrm{sca}^p\bigg(\frac{\partial J_\nu^p}{\partial n_j^\ell}\bigg)\bigg(\frac{\partial I_{\mu\nu}^k}{\partial\eta_{\mu\nu}^p}\bigg).
\end{equation}}

Equation \ref{eq:partial_dev_2} consists of a linear contribution of the intensity and a summation of nonlinear cross terms due to the background scattering. Both derivatives involving $I_{\mu\nu}^k$ depend on the scheme chosen to solve the RTE. Since we used the linear piece-wise source function scheme detailed in Sect.~\ref{part:problem}, the corresponding expressions for the derivatives are
\diar{\begin{eqnarray}
\label{eq:dIdchip}
\bigg(\frac{\partial I_{\mu\nu}^k}{\partial \chi_{\mu\nu}^\ell}\bigg) &=& \bigg(\frac{\partial I_{\mu\nu}^{k+1}}{\partial \chi_{\mu\nu}^\ell}\bigg) e^{-\Delta\tau_{\mu\nu}^k} + a_\chi^k\delta_{k\ell}+b_\chi^k\delta_{k+1 \ell},\\
\bigg(\frac{\partial I_{\mu\nu}^k}{\partial \eta_{\mu\nu}^\ell}\bigg) &=& \bigg(\frac{\partial I_{\mu\nu}^{k+1}}{\partial \eta_{\mu\nu}^\ell}\bigg)e^{-\Delta\tau_{\mu\nu}^k}+a_\eta^k\delta_{k\ell}+b_\eta^k\delta_{k+1 \ell},
\end{eqnarray}}
for outgoing rays ($\mu > 0$), and
\diar{\begin{eqnarray}
\bigg(\frac{\partial I_{\mu\nu}^k}{\partial \chi_{\mu\nu}^\ell}\bigg) &=& \bigg(\frac{\partial I_{\mu\nu}^{k-1}}{\partial \chi_{\mu\nu}^\ell}\bigg) e^{-\Delta\tau_{\mu\nu}^{k-1}} + c_\chi^k\delta_{k\ell}+d_\chi^k\delta_{k-1 \ell},\\
\label{eq:dIdetam}
\bigg(\frac{\partial I_{\mu\nu}^k}{\partial \eta_{\mu\nu}^\ell}\bigg) &=& \bigg(\frac{\partial I_{\mu\nu}^{k-1}}{\partial \eta_{\mu\nu}^\ell}\bigg)e^{-\Delta\tau_{\mu\nu}^{k-1}}+c_\eta^k\delta_{k\ell}+d_\eta^k\delta_{k-1 \ell},
\end{eqnarray}}
for ingoing rays ($\mu < 0$), where the expressions of the different involved coefficients are given in Appendix \ref{ap:der}. These coefficients can be constructed from the variables used when solving the RTE to save computational time. The boundary conditions for this scheme \diar{(Sect.~\ref{part:problem})} imply that
\diar{\begin{eqnarray}
\bigg(\frac{\partial I_{\mu\nu}^{N_z}}{\partial \chi_{\mu\nu}^\ell}\bigg) = 0, \quad \bigg(\frac{\partial I_{\mu\nu}^{N_z}}{\partial \eta_{\mu\nu}^\ell}\bigg) = 0 \quad \mu > 0 \\
\bigg(\frac{\partial I_{\mu\nu}^1}{\partial \chi_{\mu\nu}^\ell}\bigg) = 0, \quad \bigg(\frac{\partial I_{\mu\nu}^1}{\partial \eta_{\mu\nu}^\ell}\bigg) = 0 \quad \mu < 0
\end{eqnarray}}
for each value of \diar{$\ell$}. It can be shown that Eqs.~\ref{eq:dIdchip} to \ref{eq:dIdetam} define two upper ($\mu > 0$) and two lower ($\mu < 0$) triangular matrices,
\diar{\begin{equation}
\label{eq:dIdchi_mat}
\bigg(\frac{\partial I_{\mu\nu}^k}{\partial \chi_{\mu\nu}^\ell}\bigg)=\begin{cases}
                        0, & \ell < k \\
            a_\chi^k, & \ell = k \\
            (b_\chi^{\ell-1}+a_\chi^{\ell}e^{-\Delta\tau_{\mu\nu}^{\ell-1}})\prod_{i=k}^{\ell-2}e^{-\Delta\tau_{\mu\nu}^{i}} & \ell > k
                 \end{cases}
\end{equation}}
for $\mu > 0$, and
\diar{\begin{equation}
\label{eq:dIdchi_mat2}
\bigg(\frac{\partial I_{\mu\nu}^k}{\partial \chi_{\mu\nu}^\ell}\bigg)=\begin{cases}
                        0, & \ell > k \\
            c_\chi^k, & \ell = k \\
            (d_\chi^{\ell+1}+c_\chi^{\ell}e^{-\Delta\tau_{\mu\nu}^{\ell}})\prod_{i=\ell+1}^{k-1}e^{-\Delta\tau_{\mu\nu}^{i}} & \ell < k
                 \end{cases}
\end{equation}}
for $\mu < 0$. These expressions are valid for internal depth points. The matrices related to the derivatives with respect to the emissivities are obtained by using the associated coefficients. These matrices can be understood as Jacobian matrices of the intensity with respect to opacities and emissivities.

The last part to detail deals with the derivative of the mean intensity term. Through its definition, we note that this quantity involves derivatives of the intensity with respect to the population densities. In brief, the full expansion of Eq.~\ref{eq:partial_dev_2} consists of an intricate self-consistent but linear system of the derivatives of the intensity with respect to the population densities. It is possible, however, to find a solution to this system that can be written as
\diar{\begin{equation}
\label{eq:J_madness}
\bigg(\frac{\partial I_{\mu\nu}^k}{\partial n_j^\ell}\bigg) = \beta_j^\ell\bigg(\frac{\partial I_{\mu\nu}^k}{\partial \chi_{\mu\nu}^\ell}\bigg)+\gamma_j^\ell\bigg(\frac{\partial I_{\mu\nu}^k}{\partial \eta_{\mu\nu}^\ell}\bigg)+\sum_p \chi_\textrm{sca}^p r_p\bigg(\frac{\partial I_{\mu\nu}^k}{\partial \eta_{\mu\nu}^p}\bigg),
\end{equation}}
where the coefficients $r_p$ and the solution derivation can be found in Appendix \ref{ap:jac}. In practice, the background-scattering contribution to the derivatives is usually very weak and can therefore be neglected.

\subsubsection{Preconditioning \diar{of JFNK with the analytical Jacobian matrix}}
\label{part:prec_JFNK}

Preconditioning the JFNK solver is troublesome when the analytical derivation of the Jacobian matrix is used. It is expensive to compute Eq.~\ref{eq:partial_dev_2} when the interest is in the Jacobi preconditioner alone, because all off-diagonal terms need to be calculated. This issue arises from the background-scattering contribution, and we therefore have to deal with the following in order to obtain the preconditioner:
\begin{itemize}
    \item The summation in Eq.~\ref{eq:partial_dev_2} involves all cross terms ($k \neq p$ terms).
    \item The presence of the mean intensity in the scattering term also involves all cross terms (Sect.~\ref{part:analytical_jac}).
\end{itemize}
 Therefore, only an approximate Jacobi preconditioner can be used at the cost of precision and potentially convergence \diar{rate} of the Krylov solver. We give three simple solutions to overcome both problems and still provide a preconditioner that dramatically improves the convergence properties of the Krylov solver:
\begin{itemize}
    \item The very first solution consists of disregarding the background-scattering contribution. In this case, Eq.~\ref{eq:partial_dev_2} becomes for \diar{$k = \ell$}
    \begin{equation}
    \bigg(\frac{\partial I_{\mu\nu}^k}{\partial n_j^k}\bigg) = \beta_j^k\bigg(\frac{\partial I_{\mu\nu}^k}{\partial \chi_{\mu\nu}^k}\bigg)+\gamma_j^k\bigg(\frac{\partial I_{\mu\nu}^k}{\partial \eta_{\mu\nu}^k}\bigg).
    \end{equation}
    \item  The second solution consists of discarding all of the cross terms to only keep the local one. The preconditioner further reduces to a local operator when we use
    \begin{equation}
    \label{eq:approx_prec}
    \bigg(\frac{\partial I_{\mu\nu}^k}{\partial n_j^k}\bigg) = \beta_j^k\bigg(\frac{\partial I_{\mu\nu}^k}{\partial \chi_{\mu\nu}^k}\bigg)+\bigg[\gamma_j^k+\chi_\textrm{sca}^k\bigg(\frac{\partial J_\nu^k}{\partial n_j^k}\bigg)^\dagger\bigg]\bigg(\frac{\partial I_{\mu\nu}^k}{\partial \eta_{\mu\nu}^k}\bigg),
    \end{equation}
    where a previous estimate of the derivative of the mean intensity is used instead of the current one. This estimate is easily computed for the next iteration by an integration of Eq.~\ref{eq:approx_prec} over all angles. This quantity can be initialized to zero (zero radiation initial guess) or the mean intensity given by the LTE solution can be calculated. 
    \item  The last solution, which is the least constraining but requires additional operations, uses the solution given by Eq.~\ref{eq:J_madness} while only considering the local terms. The corresponding operator then has
    \begin{equation}
    \bigg(\frac{\partial I_{\mu\nu}^k}{\partial n_j^k}\bigg) = \beta_j^k\bigg(\frac{\partial I_{\mu\nu}^k}{\partial \chi_{\mu\nu}^k}\bigg)+\bigg[\gamma_j^k+\chi_\textrm{sca}^k r_k\bigg]\bigg(\frac{\partial I_{\mu\nu}^k}{\partial \eta_{\mu\nu}^k}\bigg),
    \end{equation}
    where the coefficient $r_k$ is
    \begin{equation}
    r_k = \frac{\sum_\mu \omega_\mu \bigg[\beta_j^k\bigg(\frac{\partial I_{\mu\nu}^k}{\partial \chi_{\mu\nu}^k}\bigg)+\gamma_j^k\bigg(\frac{\partial I_{\mu\nu}^k}{\partial \eta_{\mu\nu}^k}\bigg)\bigg]}{1-\sum_\mu \omega_\mu \chi_\textrm{sca}^k\bigg(\frac{\partial I_{\mu\nu}^k}{\partial \eta_{\mu\nu}^k}\bigg)}.
    \end{equation}
\end{itemize}
A preconditioner that follows one of the three solutions only requires the calculation of \diar{$\ell = k$} terms in Eq.~\ref{eq:partial_dev_2}, which are only built from the coefficients $a_\chi^k$, $a_\eta^k$, $c_\chi^k$, and $c_\eta^k$ given in Appendix \ref{ap:der}. Furthermore, it can be calculated in a comparable number of operations as the residual vector $\vec F$. \diarp{In our JFNK solvers,} we use the first solution to calculate the preconditioner \diar{since the scattering terms are negligible in Eq.~\ref{eq:J_madness} compared to the other contributions, at least with the setups we considered (Sect.~\ref{part:setup})}.


\section{Results and discussion}\label{sec:res}
We have ported a simplified version of the excellent RH code \citep{Uitenbroek_2001} to Python. \diar{The latter solves the statistical equilibrium equations using the RH92 method, but with the possibility of including partial redistribution effects of scattered photons (PRD)}. This Python version does not include PRD effects, and it is significantly slower than the C-version of RH. It is implemented using modern programming constructions, however, such as classes and inheritance, which has been extremely useful in the implementation of our proof-of-concept JFNK solver because we were able to reuse most of the atom structures, opacity calculation routines, and formal solvers of the transport equation. We used the RH92 method as a reference in order to evaluate the performance of our JFNK method. We analyzed the convergence properties of different schemes and therefore did not include Ng-acceleration in our calculations \citep{1974JChPh..61.2680N}\diar{, which is implemented in the RH code}.

\subsection{Setup}
\label{part:setup}
\begin{figure}[!ht]
    \centering
    \includegraphics[width=\hsize]{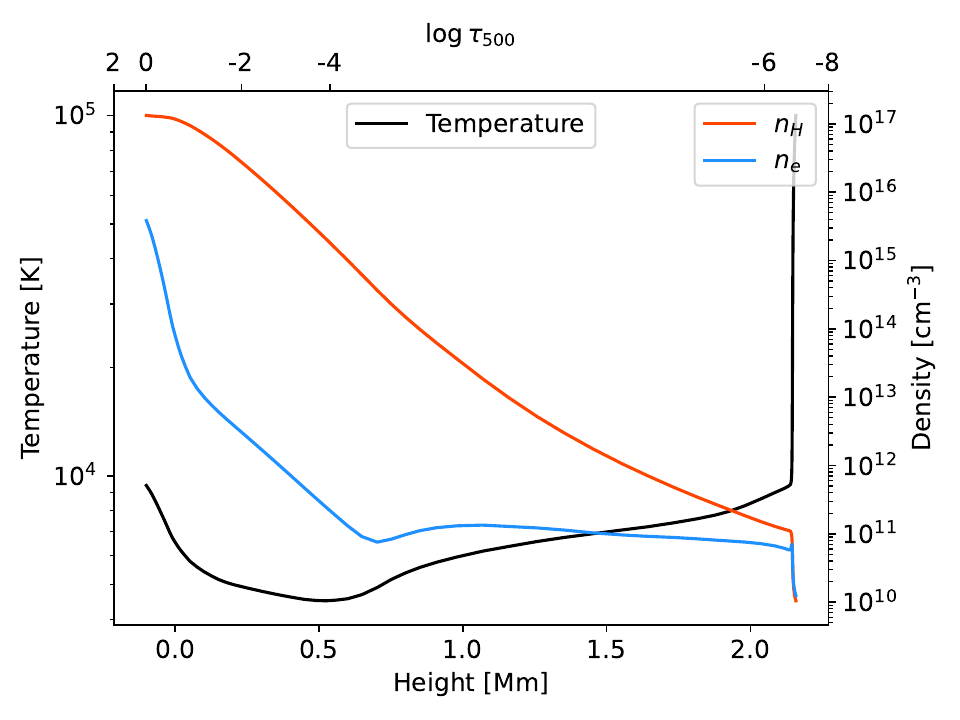}
    \caption{FAL-C atmosphere. The temperature and total hydrogen and electron densities are shown. The height dimension origin \diarp{($z = 0$)} corresponds to $\tau_{500} = 1$.}
    \label{fig:atm}
\end{figure}

All the results presented in this paper were computed using a FAL-C model atmosphere of the solar photosphere, chromosphere, and transition region \citep{1993ApJ...406..319F}, which consists of 82 depths points covering the interval $\tau_{500} = [10^{-10}, 23]$, where $\tau_{500}$ is the optical depth at $\lambda = 500$ nm. Figure \ref{fig:atm} depicts the stratifications of the gas temperature, electron density, and total hydrogen density. The atmosphere does not have a native line-of-sight velocity profile.

\diar{Three different} atomic setups were used, consisting of \diar{\ion{H}{i} and \ion{Ca}{ii}}. The different transitions are listed in table \ref{tab:setups}. Each setup also included collisional and bound-free transitions \citep{1982ApJS...48...95S,1983MNRAS.203.1269B,1985A&AS...60..425A}.
\diar{The absorption and emission profiles of each line at each location were modeled with the Voigt function, which depends on the Doppler width and the damping parameter.}
\diar{The latter includes radiative, Stark, linear Stark (in calculations with H atoms), and van der Waals broadening}. The angular \diar{integrals} were discretized using a five-point Gauss-Legendre quadrature.

\begin{table*}[!ht]
\caption{Summary of the different atom setups.}
    \centering
    \begin{tabular}{|c|c|c|}
    \hline
         Configuration & Lines wavelength [\AA] & Frequency points  \\
    \hline \hline
         \textbf{\ion{Ca}{ii}} (2 levels + continuum) & 3934 & 100 \\
         \textbf{\ion{Ca}{ii}} (5 levels + continuum) & 3934 - 3968 - 8498 - 8542 - 8662 & 100 - 100 - 40 - 40 - 80 \\
         \diar{\textbf{\ion{H}{i}}} (5 levels + continuum) & Ly($\alpha$,$\beta$,$\gamma$,$\delta$) Ba($\alpha$,$\beta$,$\gamma$) Pa($\alpha$,$\beta$) Br$\alpha$ & (150,50,20,20) (70,40,40), (20,20), 20 \\
        \hline
    \end{tabular}
    \tablefoot{ The lines of each setup are indicated as well as the number of frequency points used per line. The \ion{Ca}{ii} setup consists of the K line (three-level) and the H, K as well as the IR triplet lines (six-level). The \diar{\ion{H}{i}} setup consists of a set of Lyman, Balmer, Paschen, and Brackett lines.}
    \label{tab:setups}
\end{table*}

We performed our calculations using the Newton-Raphson, \diarp{two} JFNK (using GMRES and BiCGSTAB \diarp{, respectively}) and the RH92 methods. \diar{Both JFNK solvers systematically used the \diarp{analytical} Jacobi left preconditioner (Sect.~\ref{part:prec1} and Sect.~\ref{part:prec_JFNK}).} All solvers required an exit condition that defined a good convergence. In our case, we kept track of the residual norm $||\vec F||_\infty$ and of the population relative change norm $||\delta \vec n / \bar{\vec n}||_\infty = \frac{1}{2}||(\vec n_\textrm{new}-\vec n_\textrm{prev})/(\vec n_\textrm{new}+\vec n_\textrm{prev})||_\infty$. Unless mentioned otherwise, we assumed that a method has converged when $||\delta \vec n / \bar{\vec n}||_\infty < 10^{-4}$. This condition was sufficient most of the time, although we also imposed a minimal drop of $||\vec F||_\infty$ by two magnitudes to avoid premature exits.

\subsection{Krylov solver tolerance impact}

The Krylov solver that is internally used in the JFNK method can generally have a number of parameters that must be chosen for a given run (e.g., the size of the subspace or the convergence criteria). In the case of simple solvers such as GMRES or BiCGSTAB, this set of parameters reduces to the accuracy to which the solution is desired. This single parameter steers the behavior of the JFNK method and its convergence properties. Thus, we investigated the impact of the Krylov solver accuracy on the stability and the efficiency of the JFNK method.

\begin{figure*}[!ht]
    \centering
    \includegraphics[width=\columnwidth]{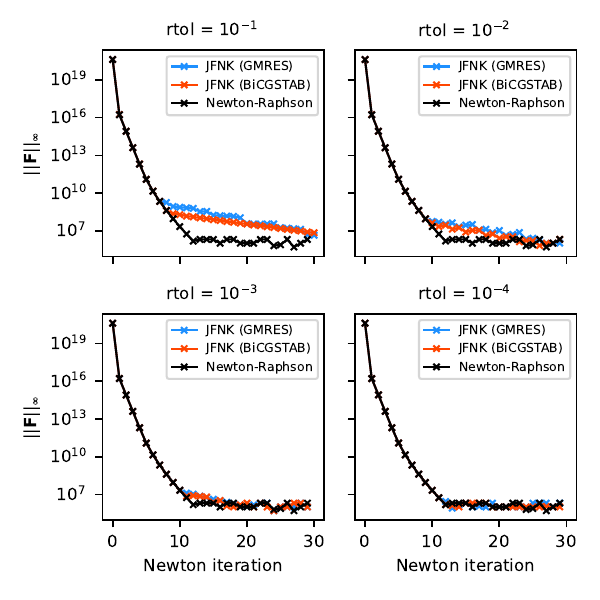}
    \includegraphics[width=\columnwidth]{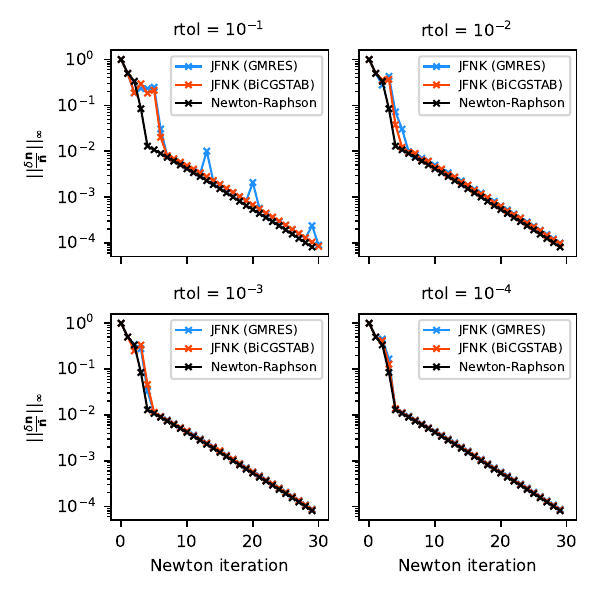}
    \caption{Comparison of the Newton\diar{-Raphson} method with JFNK routines for the \diar{six-level \ion{H}{i}} setup (zero radiation initial guess). As the Krylov solver relative tolerance $r_{\mathrm{tol}}$ becomes small, the discrepancy between the different methods reduces to truncation and round-off errors.}
    \label{fig:vs_f}
\end{figure*}

Our first test was to assess the ability of a JFNK solver to match the Newton-Raphson solution. We would expect minimal differences between the two solvers as the accuracy of the Krylov solver increases. Figure \ref{fig:vs_f} shows the convergence properties of our solvers in the case of the six-level \diar{\ion{H}{i}} atom setup. Several accuracy levels are displayed to highlight the evolution of the discrepancies between the different methods. The Newton\diar{-Raphson} solver outperforms JFNK solvers in every situation. 
%
\diar{By truncating the precision of the correction provided by the Krylov solver, each JFNK iteration becomes less accurate, usually leading to additional iterations (compared to the standard Newton-Raphson case) in order to achieve a given convergence level in the population densities. The differences between the two methods decrease when the precision of the Krylov solver is substantially increased.}
Both JFNK solvers display similar results and converge to the same solution. They show the same behavior as the Newton\diar{-Raphson} solver when $r_{\mathrm{tol}} \sim 10^{-4}$, which validates the implementation of our solvers.

\begin{figure}[!ht]
    \centering
    \includegraphics[width=\hsize]{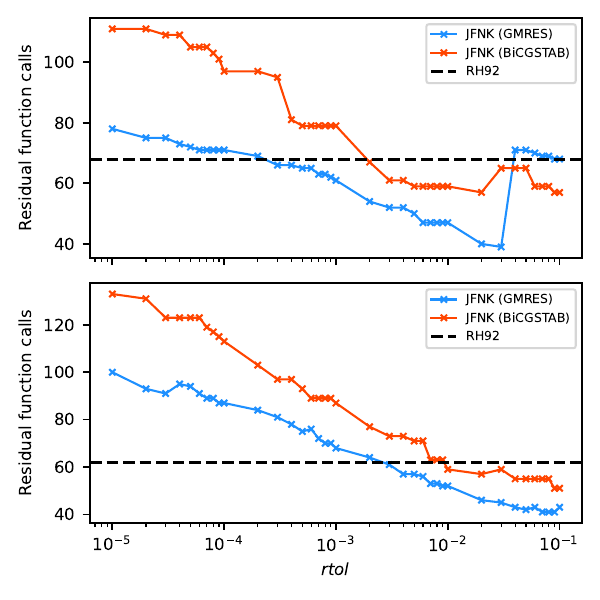}
    \caption{Residual \diar{vector} calls required for convergence as a function of the Krylov solver relative tolerance. Top: \diar{Three-level} \ion{Ca}{ii} setup. Bottom: \diar{Six-level} \ion{Ca}{ii} setup. Both setups use the LTE initial condition.}
    \label{fig:rtol_perf}
\end{figure}

A second chart is presented in Fig.~\ref{fig:rtol_perf} and directly compares our iterative solvers with the RH92 solver. In the JFNK method, we would expect a range of tolerances for which the method is optimal \diar{(with respect to the residual function calls)}:
\begin{itemize}
    \item Smaller tolerances result in more precise estimates of the inverse of the Jacobian. The JFNK method therefore requires fewer Newton\diar{-Raphson} iterations, but the overall number of calls to \diar{$\vec F$} is nonetheless higher because the additional accuracy is not effective enough in the convergence of the JFNK method. \\

    \item Higher tolerances result in coarse estimates of the inverse of the Jacobian. Even though the number of calls to \diar{$\vec F$} is reduced, Newton\diar{-Raphson} iterations usually yield poorer corrections. Therefore, the JFNK method requires additional Newton\diar{-Raphson} iterations to converge. Overall, the solver requires more calls to \diar{$\vec F$}. \\

    \item The optimal range thus consists of a trade-off between the accuracy of the inverse of the Jacobian and the calls to \diar{$\vec F$} needed to obtain them. 
\end{itemize} 

We note that overly coarse estimates of the inverse of the Jacobian can yield an unstable behavior throughout Newton\diar{-Raphson} iterations. We found a few situations in which this feature can be helpful to escape potential local minima of the residual \diar{vector}, and the method can even converge in fewer calls to \diar{$\vec F$}. These inaccurate corrections will probably not lead to the convergence of the JFNK method, however. For the sake of stability, we recommend to use a Krylov relative tolerance smaller than $\sim 10^{-2}$. 

Figure \ref{fig:rtol_perf} (top) highlights \diar{for the three-level \ion{Ca}{ii} setup} an optimal range spanning from $r_{\mathrm{tol}} \sim 3\times 10^{-4}$ to $\sim 3\times 10^{-2}$. In this range, the JFNK (GMRES) solver outperforms the RH92 solver, whereas the JFNK (BiCGSTAB) solver outperforms the latter in a much narrower range. It is also possible to witness the irregular convergence behavior of the JFNK (BiCGSTAB) solver because the corresponding curve is less smooth than that of the JFNK (GMRES) solver. In the case of a six-level setup (Fig.~\ref{fig:rtol_perf} bottom), there is no such optimal range. Instead, the number of residual \diar{vector} calls decreases almost monotonically with the Krylov solver tolerance. Both JFNK solvers outperform the RH92 solver for Krylov \diarp{relative} tolerances higher than $\sim 3\times 10^{-3}$ (GMRES) and $\sim 1\times 10^{-2}$ (BiCGSTAB). The JFNK (GMRES) solver has always proven to outperform the JFNK (BiCGSTAB) solver for most Krylov tolerances and various setups, and it outperforms the RH92 solver in a wider range of Krylov tolerances than the BiCGSTAB counterpart. The residual \diar{vector} calls from the JFNK / Newton\diar{-Raphson} solvers and the iterations of the RH92 solver are different, however. Therefore, Fig.~\ref{fig:rtol_perf} only makes sense when the two are comparable in operations or execution time, which is the case for our setups (see Sect.~\ref{part:perfs}). Thus, the Jacobi preconditioner allows the JFNK method to be more efficient than the RH92 method when it is used optimally. The preconditioner is less efficient for a setup consisting of many frequencies, however, such as the six-level \diar{\ion{Ca}{ii} or \ion{H}{i} ones}. Moreover, the lack of an optimal range of Krylov tolerances indicates that the Jacobi preconditioner is not enough for setups like this. This statement also includes every setup with strong scattering.

\subsection{Performance of the solver}
\label{part:perfs}
Table \ref{tab:exec_time} shows the average execution time per call to \diar{$\vec F$} (or equivalently, for the RH92 solver) for different setups. The pure call to \diar{$\vec F$} is always slightly faster to execute by the JFNK solvers than the RH92 solver equivalent because the RH92 solver method itself requires the computation of cross-coupling terms and the remaining rate matrix elements in order to update the population \diar{densities}. On the other hand, computing the residual vector and updating the preconditioner (JFNK) requires approximately twice the time of a pure \diar{$\vec F$} call \diar{(by a JFNK solver)}, which was expected. The preconditioner update call, while more time-consuming than the RH92 solver equivalent, is only performed once per Newton\diar{-Raphson} iteration. The main contribution to the execution time is due to the Krylov solver calls, and therefore, to pure residual vector estimates. As a result, it can be shown that the JFNK solver calls, as implemented in our proof-of-concept code, do require slightly less time on average than RH92 calls, even for extremely suboptimal Krylov tolerances.
\begin{table*}[!ht]
    \caption{Average execution time per call for the JFNK and the RH92 solvers.}
    \centering
    \begin{tabular}{|c|c|c|c|}
         \hline
     Setup & JFNK [ms / call] & JFNK (with preconditioner) [ms / call] & \text{RH92} [ms / call] \\
     \hline\hline 
     \diar{\textbf{\ion{Ca}{ii}} (2 levels + continuum)} & 55 & 121 & 65 \\
     \diar{\textbf{\ion{Ca}{ii}} (5 levels + continuum)} & 254 & 672 & 310 \\
     \diar{\textbf{\ion{H}{i}} (5 levels + continuum)} & 510 & 1160 & 631 \\
     \hline
    \end{tabular}
    
    \label{tab:exec_time}
    \tablefoot{Each call consists of the operations required to update the population \diar{densities} once (and optionally, of the preconditioner for the JFNK solvers). None of the codes was designed to be optimal. The JFNK solvers are slightly more efficient than the RH92 solver (without the preconditioner update). Runs were executed on a MacBook Pro provided with a Apple M1 Pro chip.}
\end{table*}

In the following part, we compare the quality of the solutions provided by JFNK solvers with the reference RH92 case.
The convergence condition is usually given by a sufficiently small change in the population \diar{densities}. For this purpose, we used the JFNK residual norm $||\vec F||_\infty$ as the metric because it is derived from the raw equations we attempted to solve, although the RH92 solver is not designed to minimize the raw residual norm. Moreover, the residual norm might stall when using RH92 because of the deepest layers of the atmosphere. The medium indeed becomes strongly collisional, and therefore, the radiative contribution and the changes that may occur during the solving process do not have a significant impact. Nevertheless, we disregarded these layers in the estimation of the residual norm because they are close to LTE. In the following, the residual norm is evaluated considering only the first 50 points ($z>700$~km) of our atmosphere, where the chromosphere is located. A very large error in the RH92 curve is obtained when this is neglected, and the error is mostly driven by deeper layers where LTE should hold. This behavior was absent in the \diar{JFNK} calculations. 

\begin{figure}[!ht]
    \centering
    \includegraphics[width=\hsize]{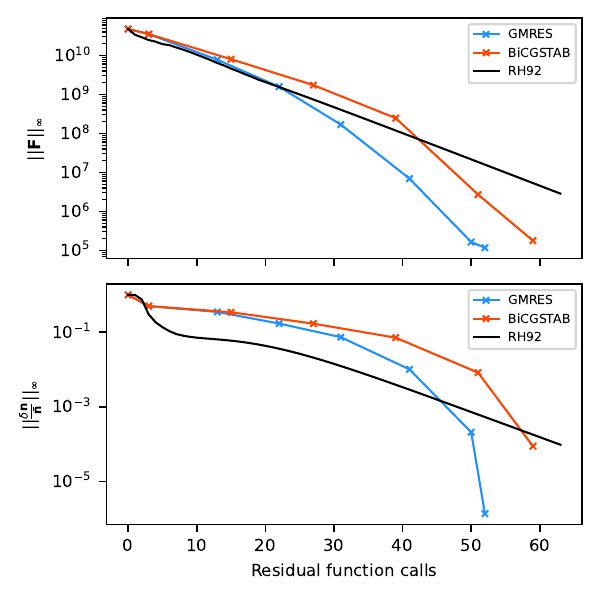}
    \caption{Residual and population change norms during the solving process of the \diar{six-level} \ion{Ca}{ii} setup. The initial population \diar{densities} are the LTE ones. Both JFNK solvers used a Krylov relative tolerance of $10^{-2}$. Each cross marker corresponds to a Newton\diar{-Raphson} iteration.}
    \label{fig:quality_CaIIall}
\end{figure}

\begin{figure}[!ht]
    \centering
    \includegraphics[width=\hsize]{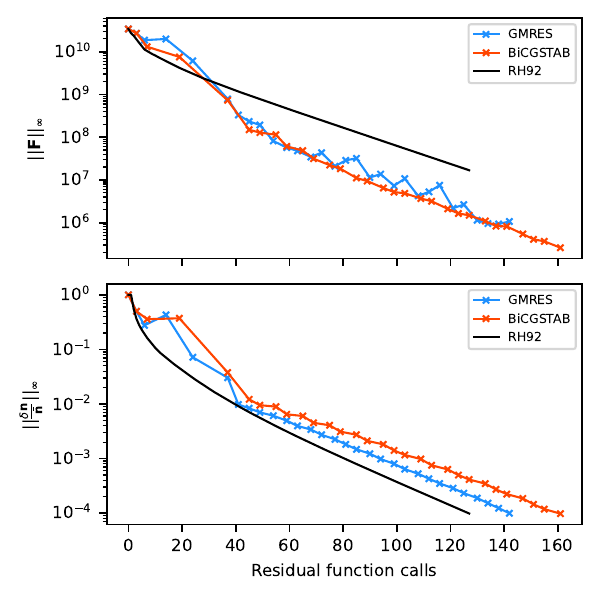}
    \caption{Residual and population change norms during the solving process of the \diar{six-level \ion{H}{i}} setup. The initial population \diar{densities} are the zero radiation initial guess ones. Both JFNK solvers used a Krylov relative tolerance of $10^{-2}$. Each cross marker corresponds to a Newton\diar{-Raphson} iteration.}
    \label{fig:quality_H_6}
\end{figure}

Figures \ref{fig:quality_CaIIall} \diar{and \ref{fig:quality_H_6}} show this clamped residual norm as well as the population change norm for the \ion{Ca}{ii} and the \ion{H}{i} setups, respectively. It is clear that the RH92 solver displays a \diar{lower convergence rate} for the largest part of the solving process. JFNK solvers, on the other hand, are outperformed at the beginning before the convergence \diar{rate surpasses that of RH92} (\diar{\ion{Ca}{ii}}) or \diar{equals the latter} (\diar{\ion{H}{i}}). As an outcome, JFNK solvers can perform better than the RH92 solver, especially when the initial guess is close enough to the solution. Moreover, the two figures show that the solution provided by the JFNK solvers is one hundred times more precise than that of RH92. We recall that the success condition is dictated by the population change \diar{norm} and not by the residual norm. 

\begin{figure}[!ht]
    \centering
    \includegraphics[width=\hsize, trim=0 0 0 5.1cm, clip]{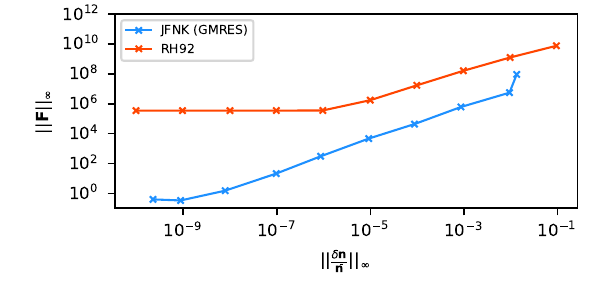}
      \caption{Residual norm of the rate equations (six-level \ion{H}{i} setup) as a function of the \diar{population change norm} for the JFNK (blue) and RH92 (red) schemes. The Krylov relative tolerance was set to $10^{-2}$. The population \diar{densities} were initialized using the zero radiation approximation. \diar{Each solver was run in order to achieve  several Newton relative tolerances in the population change norm (e.g., $10^{-1}$, $10^{-2}$). We then recorded the final residual and population change norms.}}  
    \label{fig:RH92_vs_JFNK}
\end{figure}

The size of the population \diar{change norm} is a poor criterion for convergence. Figure~\ref{fig:RH92_vs_JFNK} shows that the maximum error in the rate equations for the JFNK solver is lower than in the RH92 case for the same correction size. The JFNK solver achieves a lower absolute error in the \diar{residual norm} than RH92 for any given convergence condition set on the size of the population \diar{change norm}. This is expected because the size of the correction per iteration is affected by the efficiency of the solver: For example, in the extreme case of a traditional lambda iteration, this leads to very small corrections with a very large error in the rate equations \diar{(i.e., residual norm)}, whereas in an accelerated lambda iteration, the convergence is comparatively more efficient.


\begin{figure*}[!ht]
    \centering
    \includegraphics[width=\hsize]{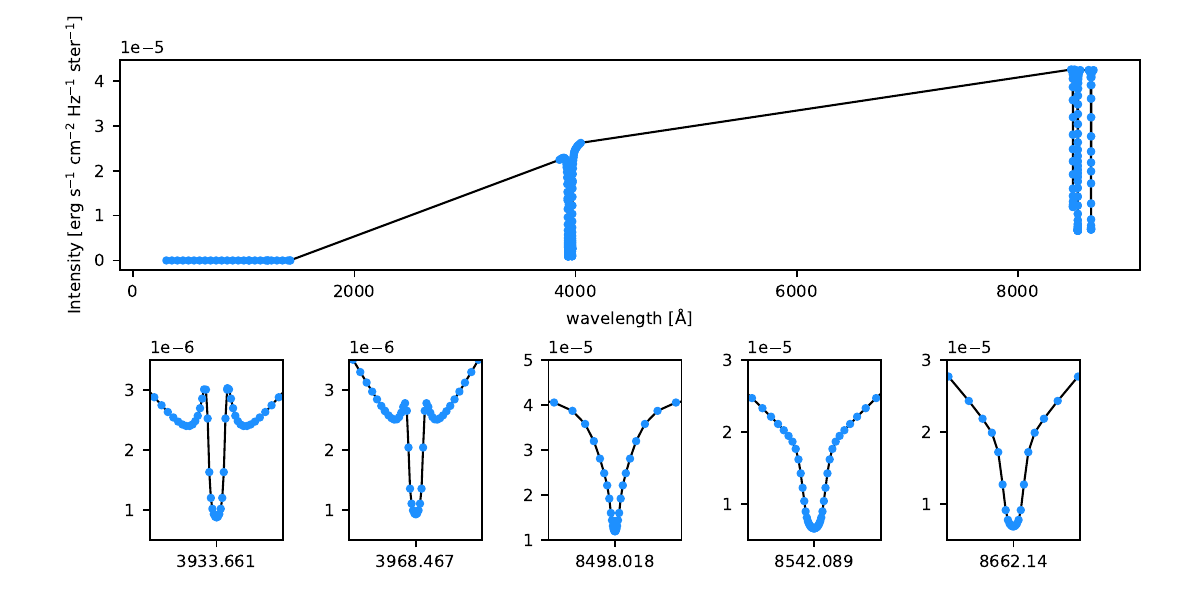}
    \caption{\ion{Ca}{ii} (six-level) spectrum. In black, we plot the output from the RH92 solver. In blue, we plot the output from the JFNK solver (GMRES) with a Krylov \diarp{relative} tolerance of $10^{-2}$.}
    \label{fig:spectrum}
\end{figure*}

\begin{figure*}[!ht]
    \centering
    \includegraphics[width=\hsize]{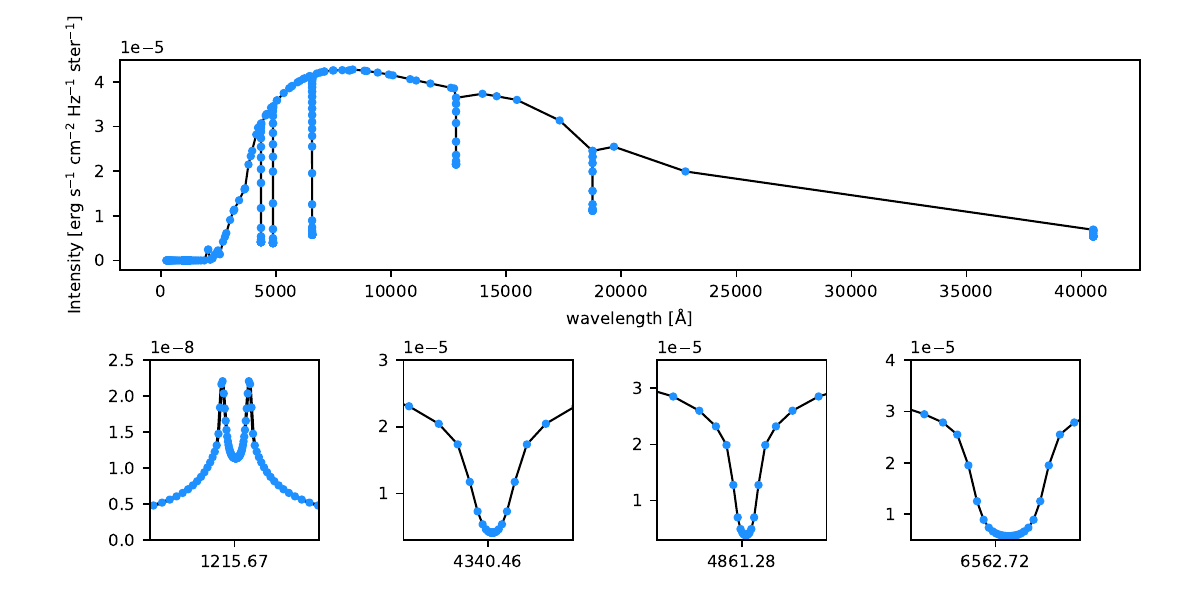}
    \caption{\diar{Six-level \ion{H}{i}} spectrum. In black, we plot the output from the RH92 solver. In blue, we plot the output from the JFNK solver (GMRES) with a Krylov \diarp{relative} tolerance of $10^{-2}$.}
    \label{fig:spectrum2}
\end{figure*}

Finally, we provide in Figs. \ref{fig:spectrum}-\ref{fig:spectrum2} the spectra of the \diar{six-level} \ion{Ca}{ii} and \diar{six-level} \ion{H}{i} setups, respectively. In both cases, the emerging spectra predicted by the RH92 and the JFNK solvers are essentially identical. We note that the additional accuracy of the solution provided by a JFNK solver does not yield noticeable changes in the emerging intensity, and we can therefore decide to terminate the solving process earlier and still output a similar result.

\subsection{Calculations with velocity gradients}
In this section, we show that the \diar{JFNK} solver can properly handle nonstatic atmospheres with velocity gradients as a function of depth. We modified the FAL-C model atmosphere by introducing a sharp velocity gradient around $z=1000$~km, corresponding to the lower chromosphere. Velocity gradients can be problematic when the velocity jump between consecutive grid cells is larger than approximately one-third of the Doppler width \citep{2013A&A...549A.126I}. Under these circumstances, the discretization of the RTE can lead to artifacts in the intensity. 

\begin{figure*}[!ht]
    \centering
    \includegraphics[width=\columnwidth]{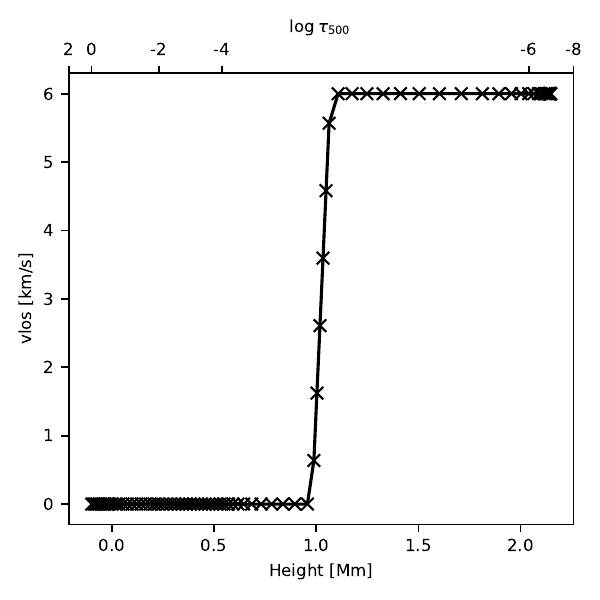}
    \includegraphics[width=\columnwidth]{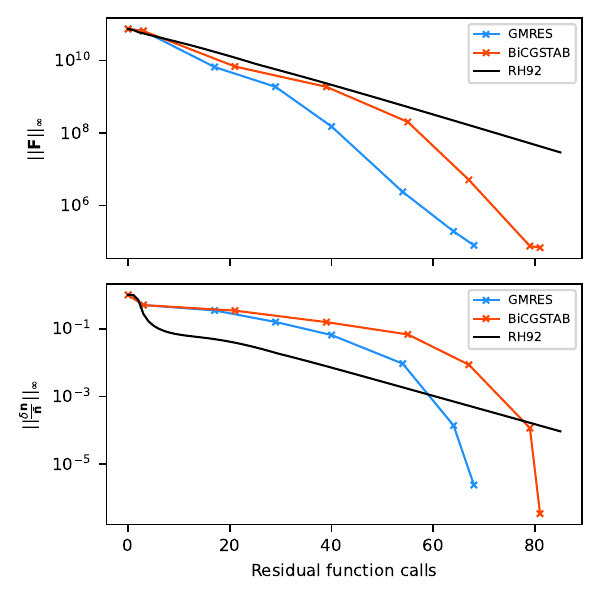}
    \includegraphics[width=\hsize]{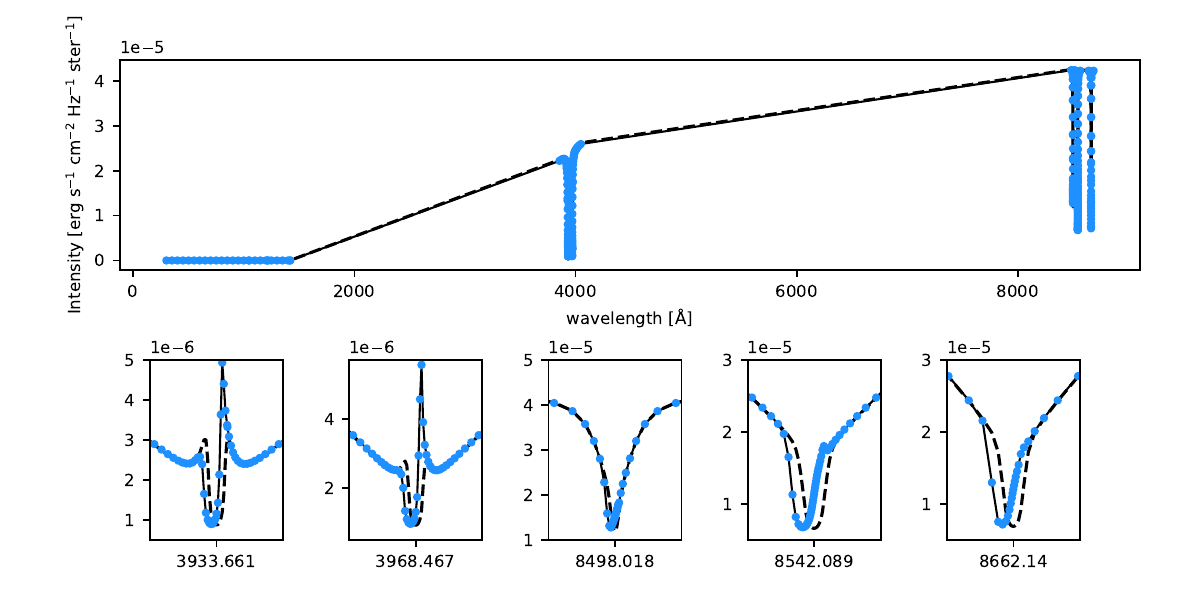}
    \caption{Velocity gradient convergence test for the different solvers. Top left panel: Line-of-sight velocity profile used for the test. Top right panel: Associated convergence plot for the \diar{six-level} \ion{Ca}{ii} setup with a Krylov \diarp{relative} tolerance of $10^{-2}$ and initial LTE population \diar{densities}. All the solvers required more iterations to converge than \diar{in the case} shown in Fig.~\ref{fig:quality_CaIIall}. Bottom panel: \diar{Converged spectrum including the velocity gradient for RH92 (black) and JFNK (GMRES) (blue). A JFNK (GMRES) velocity-free reference spectrum (dashed black) is also shown.} A blueshift clearly occurs near the line center due to the positive velocity gradient at the base of the chromosphere, resulting in very asymmetric output lines.}
    \label{fig:velgrad}
\end{figure*}

In order to avoid numerical artifacts in the calculation of the intensity, we performed a depth optimization by placing more points where the gradients in temperature, density, optical depth or line-of-sight velocity are large. All quantities were interpolated to the new depth grid by linear interpolation. The total number of depth points was kept equal to that in the original model. This method is essentially an extension of the depth-optimization included in the Multi code \citep{1986UppOR..33.....C}, which now also accounts for the presence of velocity gradients. The upper left panel in Fig.~\ref{fig:velgrad} illustrates the artificial velocity gradient represented in the optimized grid. 

When the velocity gradient is properly sampled with sufficient depth points, there is no fundamental reason why any of the algorithms would perform very differently than in the static case. Our convergence plots in Fig.~\ref{fig:velgrad} show a very similar behavior than those in Fig.~\ref{fig:quality_CaIIall} for the \ion{Ca}{ii} atom. After a few iterations, the \diar{residual norm} $||\boldsymbol{F}||_\infty$ is lower than in the RH92 curve, whereas the population change norm $||\delta \boldsymbol{n} / \boldsymbol{n}||_\infty$ \diar{is} larger. After approximately 80 iterations, the RH92 method has achieved a convergence (in the residual norm) that is similar to that of the GMRES after 50 iterations.

The emerging intensity spectrum now shows strong asymmetries around the core of all chromospheric lines, which become progressively more blueshifted by the presence of the positive velocity gradient at the base of the chromosphere. The \ion{Ca}{ii}~H\&K lines \diar{($3968~\AA$ and $3934~\AA$)} show the well-known enhancement of one of the k2 peaks (in the case of the red wing) because the blueshifted line profile in the core frequencies leaves an opacity gap in the red wing of the line where photons can escape more efficiently compared to the static case \citep{1984mrt..book..173S}. 

\subsection{Prospects}
The \diar{proposed} JFNK method can be upgraded for a better efficiency, and there are several ways of doing so. The external Newton\diar{-Raphson} update does not leave much space for improvement, but it might be interesting to implement a continuation or a line-search method to potentially reduce the number of Newton\diar{-Raphson} iterations. A nice survey of continuation methods is given in \cite{allgower_georg_1993}. A potentially simple but robust modification would be to implement a hybrid solver mixing the RH92 and the JFNK solvers. Starting with a few RH92 iterations before switching to JFNK iterations would allow this hybrid solver to avoid the usual Newton method deficiencies, as well as providing a better initial guess for the Newton-based solver. This behavior was encountered when attempting to solve the problem for instance with the \diar{six-level \ion{H}{i}} setup starting with the LTE population \diar{densities}. The performance of our JFNK solvers otherwise reduces to \diar{how efficient the Krylov solver can be,} and this is therefore \diar{dictated by} the quality of the preconditioner and the solver itself.

The Jacobi preconditioner \diar{we used} has proven to be \diar{relatively inefficient in} several \diar{of our} setups, and therefore, \diar{it could} be \diar{improved}. \diar{In our implementation, the local preconditioner is a block-diagonal matrix. When it multiplies the Jacobian, it destroys the nonlocal derivatives in the left-hand side of Eq.~\ref{eq:prec2}, and therefore, it has a similar effect as the adoption of a local approximate operator in the RH92 method.} However, we recall that the preconditioner should be kept to be easily invertible and calculable, thus leaving a narrow space for improvement. To do this, we provide two possible routes to calculate a more suitable preconditioner. The first option deals with the single-point quadrature approximation of the RTE from \cite{SCHARMER198556}. An approximation of this kind could greatly simplify the calculations of the nonlocal part of the Jacobian matrices and therefore might provide a more accurate preconditioner than the Jacobi one. The second option is more related to the JFNK formalism and is presented in \cite{adaptative} and deals with an adaptive preconditioning technique. In brief, we can take advantage of the matrix vector products calculated by a Krylov solver to iteratively update the preconditioner. It also allows us to computate a nonlocal contribution to the preconditioner.


Another upgrade that can be implemented deals with the initial guess provided to the JFNK solver. In this paper, we used two possible initial guesses, which are the LTE and the zero radiation ones. However, there might be other possibilities more suited for a Newton-based method applied to the radiative transfer problem such as the JFNK one. For instance, the population \diar{densities} can be initialized with those derived from the escape probability theory \citep[e.g.,][]{hubeny2014theory,2017ApJ...851....5J}.

\diar{We only considered 1D plane-parallel NLTE problems here. The extension to 3D geometry could be possible with some considerations. First, 3D radiative transfer codes are usually domain-decomposed for parallelization purposes \citep{2009ASPC..415...87L}, where each processor or machine only has access to the properties of the atmosphere, opacities, emissivities and population densities within one subdomain. In order to implement the inner Krylov solver, we would need to collect all population densities from all subdomains and keep the vector basis of the Krylov subspace in one manager task. The key part is the evaluation of Eq.~\ref{eq:complex}, which applies a perturbation to the population densities over the entire domain. The manager would need to propagate the relevant perturbed population densities to each subdomain, but the calculation of $\vec{J}$ can be made in the same domain-decomposed way. The cost is one additional communication from the manager to the worker tasks per Krylov iteration. At current memory standards in high-performance computing centers, this approach should be reasonable.}


\section{Conclusion}\label{sec:con}

We presented a \diar{Jacobian-free} Newton-Krylov method for solving the multilevel NLTE radiative transfer problem assuming statistical equilibrium. Our implementation showed a similar convergence as the Newton-Raphson method, without ever building the full Jacobian matrix explicitly. \diarp{As a benchmark,} we solved the NLTE problem assuming plane-parallel geometry and the FAL-C model for \diar{a three-level \ion{Ca}{II}} as well as six-level \ion{Ca}{II} and \ion{H}{I} atoms, which have been commonly used in solar physics applications. We showed that our solver can converge faster than other methods based on linearization, such as RH92. The improvement in the convergence rate depends on the atom, but it usually reaches a factor $1.5 - 2$ in the best cases. The latter is evaluated in terms of the number of formal solutions needed to converge the problem. However, we note that the \diar{JFNK} formal solutions are faster because no cross-term summations are required compared to RH92. The downside of our method is that it relies on an appropriate election of the convergence \diar{tolerance} for the Krylov inner solver. Our sensitivity study seems to indicate that an optimal performance can be attained when the tolerance is set in the range of $10^{-3} - 10^{-2}$.

In order to increment the accuracy of the Newton\diar{-Raphson} correction per iteration, we augmented the precision of the formal solver using complex numbers. This change was required given the enormous dynamic range of the atomic level population \diar{densities} from the photosphere to the transition region.

Compared to other studies that used Krylov-subspace methods to iteratively solve the linear two-level atom problem \citep[e.g.,][]{2007ApJ...659.1458H,2009ApJ...704..661A,Benedusi2021,Benedusi2022}, our method handles multilevel nonlinear problems. Because the Jacobian matrix does not need to be explicitly computed in each iteration, this method becomes particularly interesting for more complex problems, which we briefly discuss hereafter as future prospects.

The more obvious application relates to problems where partial redistribution effects of scattered photons are important. While several efficient solutions are available for the two-level atom problem \citep[e.g.,][]{1983A&A...117...83S,1995A&A...297..771P}, similar methods for multilevel problems have important limitations. For example, \citet{1995ApJ...455..376H} presented a PRD method based on the complete linearization approach of \citet{SCHARMER198556}, which does not consider overlapping active transitions. \citet{Uitenbroek_2001} overcame that limitation by using the RH92 formalism and performing two iterative cycles, separating the correction to the \diar{atomic level} population \diar{densities} and the correction to the emissivity profile. The method converges, but it requires several evaluations of the redistribution integral per iteration. The method presented in this paper shows great potential to accelerate the convergence of PRD problems because it does not require any explicit linearization of the problem.

Another extension could be the inclusion of charge conservation when H atoms are solved. The idea would be to add another conservation equation and update the electron density in each iteration because the ionization of H is usually dominant in the chromosphere. Previous studies have included these corrections, but needed to perform Newton-Raphson iterations due to the nonlinear dependences of the Saha equation and the rate equations on the electron density \citep[e.g.,][]{2007A&A...473..625L,2019PhDT.......176B}. Since we did not perform any explicit linearization of the rate equations or the transfer equation and we already used Newton\diar{-Raphson} iterations, the inclusion of charge conservation could be very efficient and relatively straightforward.

\begin{acknowledgements}
\diar{We are very thankful to the referee for his/her constructive suggestions and careful evaluation of our manuscript.}
The Institute for Solar Physics is supported by a grant for research infrastructures of national importance from the Swedish Research Council (registration number 2021-00169). 
JL acknowledges financial support from the Swedish Research council (VR, project number 2022-03535).
No animals were harmed in the making of this manuscript. 
This project has been funded by the European Union through the European Research Council (ERC) under the Horizon 2020 research and innovation program (SUNMAG, grant agreement 759548) and the Horizon Europe program (MAGHEAT, grant agreement 101088184). 
Part of our computations were enabled by resources provided by the National Academic Infrastructure for Supercomputing in Sweden (NAISS), partially funded by the Swedish Research Council through grant agreement no. 2022-06725, at the PDC Center for High Performance Computing, KTH Royal Institute of Technology (project numbers NAISS 2023/1-15 and NAISS 2024/1-14).
 \end{acknowledgements}
\bibliographystyle{aa}
\bibliography{refs}

@article{KNOLL2004357,
title = {Jacobian-free Newton–Krylov methods: a survey of approaches and applications},
journal = {Journal of Computational Physics},
volume = {193},
number = {2},
pages = {357-397},
year = {2004},
issn = {0021-9991},
doi = {https://doi.org/10.1016/j.jcp.2003.08.010},
url = {https://www.sciencedirect.com/science/article/pii/S0021999103004340},
author = {D.A. Knoll and D.E. Keyes}
}

@article{ refId0,
	author = {{Mili{\'c}, I.} and {van Noort, M.}},
	title = {Line response functions in nonlocal thermodynamic equilibrium - Isotropic case},
	DOI= "10.1051/0004-6361/201629980",
	url= "https://doi.org/10.1051/0004-6361/201629980",
	journal = {A\&A},
	year = 2017,
	volume = 601,
	pages = "A100",
}

@ARTICLE{2018A&A...617A..24M,
       author = {{Mili{\'c}}, I. and {van Noort}, M.},
        title = "{Spectropolarimetric NLTE inversion code SNAPI}",
      journal = {\aap},
         year = 2018,
        month = sep,
       volume = {617},
          eid = {A24},
        pages = {A24},
          doi = {10.1051/0004-6361/201833382},
archivePrefix = {arXiv},
       eprint = {1806.08134},
 primaryClass = {astro-ph.SR},
       adsurl = {https://ui.adsabs.harvard.edu/abs/2018A&A...617A..24M}
}

@ARTICLE{1992A&A...262..209R,
       author = {{Rybicki}, G.~B. and {Hummer}, D.~G.},
        title = "{An accelerated lambda iteration method for multilevel radiative transfer. II. Overlapping transitions with full continuum.}",
      journal = {\aap},
         year = 1992,
        month = aug,
       volume = {262},
        pages = {209-215},
       adsurl = {https://ui.adsabs.harvard.edu/abs/1992A&A...262..209R}
}

@article{KAN2022113732,
title = {Extension of complex step finite difference method to Jacobian-free Newton–Krylov​ method},
journal = {Journal of Computational and Applied Mathematics},
volume = {399},
pages = {113732},
year = {2022},
issn = {0377-0427},
doi = {https://doi.org/10.1016/j.cam.2021.113732},
url = {https://www.sciencedirect.com/science/article/pii/S037704272100354X},
author = {Ziyun Kan and Ningning Song and Haijun Peng and Biaosong Chen}
}

@article{Uitenbroek_2001,
doi = {10.1086/321659},
url = {https://dx.doi.org/10.1086/321659},
year = {2001},
month = {aug},
publisher = {},
volume = {557},
number = {1},
pages = {389},
author = {H. Uitenbroek},
title = {Multilevel Radiative Transfer with Partial Frequency Redistribution},
journal = {The Astrophysical Journal}
}

@book{doi:10.1137/1.9781611971200,
author = {Dennis, J. E. and Schnabel, Robert B.},
title = {Numerical Methods for Unconstrained Optimization and Nonlinear Equations},
publisher = {Society for Industrial and Applied Mathematics},
year = {1996},
doi = {10.1137/1.9781611971200},
address = {},
edition   = {},
URL = {https://epubs.siam.org/doi/abs/10.1137/1.9781611971200},
eprint = {https://epubs.siam.org/doi/pdf/10.1137/1.9781611971200}
}

@article{doi:10.1137/0907058,
author = {Saad, Youcef and Schultz, Martin H.},
title = {GMRES: A Generalized Minimal Residual Algorithm for Solving Nonsymmetric Linear Systems},
journal = {SIAM Journal on Scientific and Statistical Computing},
volume = {7},
number = {3},
pages = {856-869},
year = {1986},
doi = {10.1137/0907058},

URL = { 
    
        https://doi.org/10.1137/0907058
    
    

},
eprint = { 
    
        https://doi.org/10.1137/0907058
    
    

}
}

@article{doi:10.1137/0712047,
author = {Paige, C. C. and Saunders, M. A.},
title = {Solution of Sparse Indefinite Systems of Linear Equations},
journal = {SIAM Journal on Numerical Analysis},
volume = {12},
number = {4},
pages = {617-629},
year = {1975},
doi = {10.1137/0712047},

URL = { 
    
        https://doi.org/10.1137/0712047
    
    

},
eprint = { 
    
        https://doi.org/10.1137/0712047
    
    

}
}

@article{doi:10.1137/0913035,
author = {van der Vorst, H. A.},
title = {Bi-CGSTAB: A Fast and Smoothly Converging Variant of Bi-CG for the Solution of Nonsymmetric Linear Systems},
journal = {SIAM Journal on Scientific and Statistical Computing},
volume = {13},
number = {2},
pages = {631-644},
year = {1992},
doi = {10.1137/0913035},

URL = { 
    
        https://doi.org/10.1137/0913035
    
    

},
eprint = { 
    
        https://doi.org/10.1137/0913035
        }
}

@article{adaptative,
author = {Chen, Ying and Shen, Chen},
year = {2006},
month = {09},
pages = {1096 - 1103},
title = {A Jacobian-Free Newton-GMRES(m) Method with Adaptive Preconditioner and Its Application for Power Flow Calculations},
volume = {21},
journal = {Power Systems, IEEE Transactions on},
doi = {10.1109/TPWRS.2006.876696}
}

@article{SCHARMER198556,
title = {A new approach to multi-level non-LTE radiative transfer problems},
journal = {Journal of Computational Physics},
volume = {59},
number = {1},
pages = {56-80},
year = {1985},
issn = {0021-9991},
doi = {https://doi.org/10.1016/0021-9991(85)90107-X},
url = {https://www.sciencedirect.com/science/article/pii/002199918590107X},
author = {G.B Scharmer and M Carlsson}
}

@book{hubeny2014theory,
  title={Theory of Stellar Atmospheres: An Introduction to Astrophysical Non-equilibrium Quantitative Spectroscopic Analysis},
  author={Hubeny, I. and Mihalas, D.},
  isbn={9780691163291},
  lccn={2014006308},
  series={Princeton Series in Astrophysics},
  url={https://books.google.se/books?id=TmuYDwAAQBAJ},
  year={2014},
  publisher={Princeton University Press}
}

@ARTICLE{1973ApJ...185..621C,
       author = {{Cannon}, C.~J.},
        title = "{Frequency-Quadrature Perturbations in Radiative-Transfer Theory}",
      journal = {\apj},
         year = 1973,
        month = oct,
       volume = {185},
        pages = {621-630},
          doi = {10.1086/152442},
       adsurl = {https://ui.adsabs.harvard.edu/abs/1973ApJ...185..621C}
}

@ARTICLE{2012A&A...543A.109L,
       author = {{Leenaarts}, J. and {Pereira}, T. and {Uitenbroek}, H.},
        title = "{Fast approximation of angle-dependent partial redistribution in moving atmospheres}",
      journal = {\aap},
         year = 2012,
        month = jul,
       volume = {543},
          eid = {A109},
        pages = {A109},
          doi = {10.1051/0004-6361/201219394},
archivePrefix = {arXiv},
       eprint = {1205.5110},
 primaryClass = {astro-ph.SR},
       adsurl = {https://ui.adsabs.harvard.edu/abs/2012A&A...543A.109L}
}

@INPROCEEDINGS{2009ASPC..415...87L,
       author = {{Leenaarts}, J. and {Carlsson}, M.},
        title = "{MULTI3D: A Domain-Decomposed 3D Radiative Transfer Code}",
    booktitle = {The Second Hinode Science Meeting: Beyond Discovery-Toward Understanding},
         year = 2009,
       editor = {{Lites}, B. and {Cheung}, M. and {Magara}, T. and {Mariska}, J. and {Reeves}, K.},
       series = {Astronomical Society of the Pacific Conference Series},
       volume = {415},
        month = dec,
        pages = {87},
       adsurl = {https://ui.adsabs.harvard.edu/abs/2009ASPC..415...87L}
}

@ARTICLE{2018A&A...615A.139A,
       author = {{Amarsi}, A.~M. and {Nordlander}, T. and {Barklem}, P.~S. and {Asplund}, M. and {Collet}, R. and {Lind}, K.},
        title = "{Effective temperature determinations of late-type stars based on 3D non-LTE Balmer line formation}",
      journal = {\aap},
         year = 2018,
        month = jul,
       volume = {615},
          eid = {A139},
        pages = {A139},
          doi = {10.1051/0004-6361/201732546},
archivePrefix = {arXiv},
       eprint = {1804.02305},
 primaryClass = {astro-ph.SR},
       adsurl = {https://ui.adsabs.harvard.edu/abs/2018A&A...615A.139A}
}

@ARTICLE{1986JQSRT..35..431O,
       author = {{Olson}, G.~L. and {Auer}, L.~H. and {Buchler}, J.~R.},
        title = "{A rapidly convergent iterative solution of the non-LTE radiation transfer problem.}",
      journal = {\jqsrt},
         year = 1986,
        month = jun,
       volume = {35},
        pages = {431-442},
          doi = {10.1016/0022-4073(86)90030-0},
       adsurl = {https://ui.adsabs.harvard.edu/abs/1986JQSRT..35..431O}
}

@ARTICLE{1981ApJ...249..720S,
       author = {{Scharmer}, G.~B.},
        title = "{Solutions to radiative transfer problems using approximate lambda operators}",
      journal = {\apj},
         year = 1981,
        month = oct,
       volume = {249},
        pages = {720-730},
          doi = {10.1086/159333},
       adsurl = {https://ui.adsabs.harvard.edu/abs/1981ApJ...249..720S}
}

@ARTICLE{1982StoOR..19.....S,
       author = {{Scharmer}, G.~B. and {Nordlund}, {\r{A}}.},
        title = "{DQPT: a computer program for solving non-LTE problems for two-level atoms in one-dimensional semi-infinite media with velocity fields.}",
      journal = {Stockholms Observatoriums Reports},
         year = 1982,
        month = jan,
       volume = {19},
       adsurl = {https://ui.adsabs.harvard.edu/abs/1982StoOR..19.....S}
}

@ARTICLE{1921MNRAS..81..361M,
       author = {{Milne}, E.~A.},
        title = "{Radiative equilibrium in the outer layers of a star}",
      journal = {\mnras},
         year = 1921,
        month = mar,
       volume = {81},
        pages = {361-375},
          doi = {10.1093/mnras/81.5.361},
       adsurl = {https://ui.adsabs.harvard.edu/abs/1921MNRAS..81..361M}
}

@BOOK{1926ics..book.....E,
       author = {{Eddington}, A.~S.},
        title = "{The Internal Constitution of the Stars}",
         year = 1926,
       adsurl = {https://ui.adsabs.harvard.edu/abs/1926ics..book.....E}
}

@ARTICLE{1943AnAp....6..113B,
       author = {{Barbier}, Daniel},
        title = "{Sur la th{\'e}orie du spectre continu des {\'e}toiles}",
      journal = {Annales d'Astrophysique},
         year = 1943,
        month = jan,
       volume = {6},
        pages = {113},
       adsurl = {https://ui.adsabs.harvard.edu/abs/1943AnAp....6..113B}
}

@ARTICLE{2015A&A...574A...3P,
       author = {{Pereira}, Tiago M.~D. and {Uitenbroek}, Han},
        title = "{RH 1.5D: a massively parallel code for multi-level radiative transfer with partial frequency redistribution and Zeeman polarisation}",
      journal = {\aap},
         year = 2015,
        month = feb,
       volume = {574},
          eid = {A3},
        pages = {A3},
          doi = {10.1051/0004-6361/201424785},
archivePrefix = {arXiv},
       eprint = {1411.1079},
 primaryClass = {astro-ph.SR},
       adsurl = {https://ui.adsabs.harvard.edu/abs/2015A&A...574A...3P}
}

@ARTICLE{1986UppOR..33.....C,
       author = {{Carlsson}, M.},
        title = "{A computer program for solving multi-level non-LTE radiative transferproblems in moving or static atmospheres.}",
      journal = {Uppsala Astronomical Observatory Reports},
         year = 1986,
        month = jan,
       volume = {33},
       adsurl = {https://ui.adsabs.harvard.edu/abs/1986UppOR..33.....C}
}

@ARTICLE{2015A&A...577A...7S,
       author = {{Socas-Navarro}, H. and {de la Cruz Rodr{\'\i}guez}, J. and {Asensio Ramos}, A. and {Trujillo Bueno}, J. and {Ruiz Cobo}, B.},
        title = "{An open-source, massively parallel code for non-LTE synthesis and inversion of spectral lines and Zeeman-induced Stokes profiles}",
      journal = {\aap},
         year = 2015,
        month = may,
       volume = {577},
          eid = {A7},
        pages = {A7},
          doi = {10.1051/0004-6361/201424860},
archivePrefix = {arXiv},
       eprint = {1408.6101},
 primaryClass = {astro-ph.SR},
       adsurl = {https://ui.adsabs.harvard.edu/abs/2015A&A...577A...7S}
}

@book{Raphson2021-RAPAAU-2,
	author = {Joseph Raphson},
	editor = {},
	title = {Analysis Aequationum Universalis Seu Ad Aequationes Algebraicas Resolvendas Methodus Generalis, \& Expedita, Ex Nova Infinitarum Serierum Methodo, Deducta Ac Demonstrata},
	year = {1690}
}

@book{newton1736method,
  title={The Method of Fluxions and Infinite Series: With Its Application to the Geometry of Curve-lines. By ... Sir Isaac Newton, ... Translated from the Author's Latin Original Not Yet Made Publick. To which is Subjoin'd, a Perpetual Comment Upon the Whole Work, ... By John Colson, ...},
  author={Newton, I.},
  url={https://books.google.se/books?id=WyQOAAAAQAAJ},
  year={1736},
  publisher={Henry Woodfall; and sold by John Nourse}
}

@article{krylov1931numerical,
  title={On the numerical solution of the equation by which the frequency of small oscillations is determined in technical problems},
  author={Krylov, Aleksei N},
  journal={Izv. Akad. Nauk SSSR Ser. Fiz.-Mat},
  volume={4},
  pages={491--539},
  year={1931}
}

@ARTICLE{1969ApJ...158..641A,
       author = {{Auer}, Lawrence H. and {Mihalas}, Dimitri},
        title = "{Non-Lte Model Atmospheres. III. a Complete-Linearization Method}",
      journal = {\apj},
         year = 1969,
        month = nov,
       volume = {158},
        pages = {641},
          doi = {10.1086/150226},
       adsurl = {https://ui.adsabs.harvard.edu/abs/1969ApJ...158..641A}
}

@ARTICLE{1993ApJ...406..319F,
       author = {{Fontenla}, J.~M. and {Avrett}, E.~H. and {Loeser}, R.},
        title = "{Energy Balance in the Solar Transition Region. III. Helium Emission in Hydrostatic, Constant-Abundance Models with Diffusion}",
      journal = {\apj},
         year = 1993,
        month = mar,
       volume = {406},
        pages = {319},
          doi = {10.1086/172443},
       adsurl = {https://ui.adsabs.harvard.edu/abs/1993ApJ...406..319F}
}

@ARTICLE{1987JQSRT..38..325O,
       author = {{Olson}, G.~L. and {Kunasz}, P.~B.},
        title = "{Short characteristic solution of the non-LTE transfer problem by operator perturbation. I. The one-dimensional planar slab.}",
      journal = {\jqsrt},
         year = 1987,
        month = jan,
       volume = {38},
       number = {5},
        pages = {325-336},
          doi = {10.1016/0022-4073(87)90027-6},
       adsurl = {https://ui.adsabs.harvard.edu/abs/1987JQSRT..38..325O}
}

@ARTICLE{1974JChPh..61.2680N,
       author = {{Ng}, K. -C.},
        title = "{Hypernetted chain solutions for the classical one-component plasma up to {\ensuremath{\Gamma}}=7000}",
      journal = {\jcp},
         year = 1974,
        month = oct,
       volume = {61},
       number = {7},
        pages = {2680-2689},
          doi = {10.1063/1.1682399},
       adsurl = {https://ui.adsabs.harvard.edu/abs/1974JChPh..61.2680N}
}

@article{allgower_georg_1993, title={Continuation and path following}, volume={2}, DOI={10.1017/S0962492900002336}, journal={Acta Numerica}, publisher={Cambridge University Press}, author={Allgower, Eugene L. and Georg, Kurt}, year={1993}, pages={1–64}}

@ARTICLE{1982ApJS...48...95S,
       author = {{Shull}, J.~M. and {van Steenberg}, M.},
        title = "{The ionization equilibrium of astrophysically abundant elements.}",
      journal = {\apjs},
         year = 1982,
        month = jan,
       volume = {48},
        pages = {95-107},
          doi = {10.1086/190769},
       adsurl = {https://ui.adsabs.harvard.edu/abs/1982ApJS...48...95S}
}

@ARTICLE{1983MNRAS.203.1269B,
       author = {{Burgess}, A. and {Chidichimo}, M.~C.},
        title = "{Electron impact ionization of complex ions}",
      journal = {\mnras},
         year = 1983,
        month = jun,
       volume = {203},
        pages = {1269-1280},
          doi = {10.1093/mnras/203.4.1269},
       adsurl = {https://ui.adsabs.harvard.edu/abs/1983MNRAS.203.1269B}
}

@ARTICLE{1985A&AS...60..425A,
       author = {{Arnaud}, M. and {Rothenflug}, R.},
        title = "{An Updated Evaluation of Recombination and Ionization Rates}",
      journal = {\aaps},
         year = 1985,
        month = jun,
       volume = {60},
        pages = {425},
       adsurl = {https://ui.adsabs.harvard.edu/abs/1985A&AS...60..425A}
}

@INPROCEEDINGS{1972lfpm.conf..145R,
       author = {{Rybicki}, G.~B.},
        title = "{A Novel Approach to the Solution of Multilevel Transfer Problems}",
    booktitle = {Line Formation in the Presence of Magnetic Fields},
         year = 1972,
        month = jan,
        pages = {145},
       adsurl = {https://ui.adsabs.harvard.edu/abs/1972lfpm.conf..145R}
}

@ARTICLE{2021ApJ...917...14O,
       author = {{Osborne}, Christopher M.~J. and {Mili{\'c}}, Ivan},
        title = "{The Lightweaver Framework for Nonlocal Thermal Equilibrium Radiative Transfer in Python}",
      journal = {\apj},
         year = 2021,
        month = aug,
       volume = {917},
       number = {1},
          eid = {14},
        pages = {14},
          doi = {10.3847/1538-4357/ac02be},
archivePrefix = {arXiv},
       eprint = {2107.00475},
 primaryClass = {astro-ph.IM},
       adsurl = {https://ui.adsabs.harvard.edu/abs/2021ApJ...917...14O}
}

@ARTICLE{2017ApJ...851....5J,
       author = {{Judge}, Philip G.},
        title = "{Efficient Radiative Transfer for Dynamically Evolving Stratified Atmospheres}",
      journal = {\apj},
         year = 2017,
        month = dec,
       volume = {851},
       number = {1},
          eid = {5},
        pages = {5},
          doi = {10.3847/1538-4357/aa96a9},
       adsurl = {https://ui.adsabs.harvard.edu/abs/2017ApJ...851....5J}
}

@ARTICLE{2009ApJ...704..661A,
       author = {{Anusha}, L.~S. and {Nagendra}, K.~N. and {Paletou}, F. and {L{\'e}ger}, L.},
        title = "{Preconditioned Bi-conjugate Gradient Method for Radiative Transfer in Spherical Media}",
      journal = {\apj},
         year = 2009,
        month = oct,
       volume = {704},
       number = {1},
        pages = {661-671},
          doi = {10.1088/0004-637X/704/1/661},
archivePrefix = {arXiv},
       eprint = {0906.2926},
 primaryClass = {astro-ph.SR},
       adsurl = {https://ui.adsabs.harvard.edu/abs/2009ApJ...704..661A}
}

@ARTICLE{2007ApJ...659.1458H,
       author = {{Hubeny}, Ivan and {Burrows}, Adam},
        title = "{A New Algorithm for Two-Dimensional Transport for Astrophysical Simulations. I. General Formulation and Tests for the One-Dimensional Spherical Case}",
      journal = {\apj},
         year = 2007,
        month = apr,
       volume = {659},
       number = {2},
        pages = {1458-1487},
          doi = {10.1086/512179},
archivePrefix = {arXiv},
       eprint = {astro-ph/0609049},
 primaryClass = {astro-ph},
       adsurl = {https://ui.adsabs.harvard.edu/abs/2007ApJ...659.1458H}
}

@ARTICLE{1983A&A...117...83S,
       author = {{Scharmer}, G.~B.},
        title = "{A Linearization Method for Solving Partial Redistribution Problems}",
      journal = {\aap},
         year = 1983,
        month = jan,
       volume = {117},
        pages = {83},
       adsurl = {https://ui.adsabs.harvard.edu/abs/1983A&A...117...83S}
}

@ARTICLE{1995ApJ...455..376H,
       author = {{Hubeny}, I. and {Lites}, B.~W.},
        title = "{Partial Redistribution in Multilevel Atoms. I. Method and Application to the Solar Hydrogen Line Formation}",
      journal = {\apj},
         year = 1995,
        month = dec,
       volume = {455},
        pages = {376},
          doi = {10.1086/176584},
       adsurl = {https://ui.adsabs.harvard.edu/abs/1995ApJ...455..376H}
}

@ARTICLE{1995A&A...297..771P,
       author = {{Paletou}, F. and {Auer}, L.~H.},
        title = "{A new approximate operator method for partial frequency redistribution problems.}",
      journal = {\aap},
         year = 1995,
        month = may,
       volume = {297},
        pages = {771},
       adsurl = {https://ui.adsabs.harvard.edu/abs/1995A&A...297..771P}
}

@PHDTHESIS{2019PhDT.......176B,
       author = {{Bj{\o}rgen}, Johan Pires},
        title = "{The synthetic chromosphere: Results and techniques with a numerical approach}",
       school = {Stockholm University},
         year = 2019,
        month = jan,
       adsurl = {https://ui.adsabs.harvard.edu/abs/2019PhDT.......176B}
}

@ARTICLE{2007A&A...473..625L,
       author = {{Leenaarts}, J. and {Carlsson}, M. and {Hansteen}, V. and {Rutten}, R.~J.},
        title = "{Non-equilibrium hydrogen ionization in 2D simulations of the solar atmosphere}",
      journal = {\aap},
         year = 2007,
        month = oct,
       volume = {473},
       number = {2},
        pages = {625-632},
          doi = {10.1051/0004-6361:20078161},
archivePrefix = {arXiv},
       eprint = {0709.3751},
 primaryClass = {astro-ph},
       adsurl = {https://ui.adsabs.harvard.edu/abs/2007A&A...473..625L}
}

@ARTICLE{2013A&A...549A.126I,
       author = {{Ibgui}, L. and {Hubeny}, I. and {Lanz}, T. and {Stehl{\'e}}, C.},
        title = "{IRIS: a generic three-dimensional radiative transfer code}",
      journal = {\aap},
         year = 2013,
        month = jan,
       volume = {549},
          eid = {A126},
        pages = {A126},
          doi = {10.1051/0004-6361/201220468},
archivePrefix = {arXiv},
       eprint = {1211.4870},
 primaryClass = {astro-ph.IM},
       adsurl = {https://ui.adsabs.harvard.edu/abs/2013A&A...549A.126I}
}

@INCOLLECTION{1984mrt..book..173S,
       author = {{Scharmer}, G.~B.},
        title = "{Accurate solutions to non-LTE problems using approximate Lambda operators.}",
    booktitle = {Methods in Radiative Transfer},
         year = 1984,
        pages = {173-210},
       adsurl = {https://ui.adsabs.harvard.edu/abs/1984mrt..book..173S}
}

@ARTICLE{2017A&A...597A..46S,
       author = {{Sukhorukov}, Andrii V. and {Leenaarts}, Jorrit},
        title = "{Partial redistribution in 3D non-LTE radiative transfer in solar-atmosphere models}",
      journal = {\aap},
         year = 2017,
        month = jan,
       volume = {597},
          eid = {A46},
        pages = {A46},
          doi = {10.1051/0004-6361/201629086},
archivePrefix = {arXiv},
       eprint = {1606.05180},
 primaryClass = {astro-ph.SR},
       adsurl = {https://ui.adsabs.harvard.edu/abs/2017A&A...597A..46S}
}

@BOOK{2002nrca.book.....P,
       author = {{Press}, William H. and {Teukolsky}, Saul A. and {Vetterling}, William T. and {Flannery}, Brian P.},
        title = "{Numerical recipes in C++ : the art of scientific computing}",
         year = 2002,
       adsurl = {https://ui.adsabs.harvard.edu/abs/2002nrca.book.....P}
}

@ARTICLE{2024A&A...682A..68J,
       author = {{Janett}, Gioele and {Benedusi}, Pietro and {Riva}, Fabio},
        title = "{Numerical solutions to linear transfer problems of polarized radiation. IV. Efficient preconditioning in a physics-based framework}",
      journal = {\aap},
         year = 2024,
        month = feb,
       volume = {682},
          eid = {A68},
        pages = {A68},
          doi = {10.1051/0004-6361/202348048},
       adsurl = {https://ui.adsabs.harvard.edu/abs/2024A&A...682A..68J}
}

@book{Martins_Ning_2021, place={Cambridge}, title={Engineering Design Optimization}, publisher={Cambridge University Press}, author={Martins, Joaquim R. R. A. and Ning, Andrew}, year={2021}}

@article{Benedusi2021,
       author = {{Benedusi}, Pietro and {Janett}, Gioele and {Belluzzi}, Luca and {Krause}, Rolf},
        title = "{Numerical solutions to linear transfer problems of polarized radiation. II. Krylov methods and matrix-free implementation}",
      journal = {\aap},
         year = 2021,
        month = nov,
       volume = {655},
          eid = {A88},
        pages = {A88},
          doi = {10.1051/0004-6361/202141238},
archivePrefix = {arXiv},
       eprint = {2110.11873},
 primaryClass = {math.NA},
       adsurl = {https://ui.adsabs.harvard.edu/abs/2021A&A...655A..88B}
}

@ARTICLE{Benedusi2022,
   author = {{Benedusi}, P. and {Janett}, G. and {Riva}, G. and {Belluzzi}, L. and {Krause}, R.},
    title = "{Numerical solutions to linear transfer problems of polarized radiation.
III. Parallel preconditioned Krylov solver tailored for modeling PRD effects}",
  journal={\aap},
    volume={664},
  pages={A197},
 year = 2022
}

\begin{appendix}

\onecolumn \section{Discretization coefficients}
\subsection{Piecewise linear RTE}\label{ap:piece}
\begin{equation}
a^k = 1-\frac{1-e^{-\Delta\tau_{\mu\nu}^k}}{\Delta\tau_{\mu\nu}^k} \quad b^k = \frac{1-e^{-\Delta\tau_{\mu\nu}^k}}{\Delta\tau_{\mu\nu}^k}-e^{-\Delta\tau_{\mu\nu}^k}
\end{equation}
\subsection{Derivatives of the intensity}\label{ap:der}
For outgoing rays ($\mu > 0$),
\begin{eqnarray}
a_\chi^k &=& -\frac{S_{\mu\nu}^k}{\chi_{\mu\nu}^k}a^k+\frac{|z_{k+1}-z_k|}{2\mu}\bigg[e^{-\Delta\tau_{\mu\nu}^k}(S_{\mu\nu}^{k+1}-I_{\mu\nu}^{k+1})-\frac{S_{\mu\nu}^{k+1}-S_{\mu\nu}^k}{\Delta\tau_{\mu\nu}^k}b^k\bigg]\\
b_\chi^k &=& -\frac{S_{\mu\nu}^{k+1}}{\chi_{\mu\nu}^{k+1}}b^k+\frac{|z_{k+1}-z_k|}{2\mu}\bigg[e^{-\Delta\tau_{\mu\nu}^k}(S_{\mu\nu}^{k+1}-I_{\mu\nu}^{k+1})-\frac{S_{\mu\nu}^{k+1}-S_{\mu\nu}^k}{\Delta\tau_{\mu\nu}^k}b^k\bigg] \\
a_\eta^k &=& \frac{S_{\mu\nu}^k}{\eta_{\mu\nu}^k}a^k\\
b_\eta^k &=& \frac{S_{\mu\nu}^{k+1}}{\eta_{\mu\nu}^{k+1}}b^k
\end{eqnarray}
when $k < N_z-1;$ otherwise, one can set the coefficients to zero. For ingoing rays ($\mu < 0$),\begin{eqnarray}
c_\chi^k &=& -\frac{S_{\mu\nu}^k}{\chi_{\mu\nu}^k}a^{k-1}+\frac{|z_{k-1}-z_k|}{2\mu}\bigg[e^{-\Delta\tau_{\mu\nu}^{k-1}}(S_{\mu\nu}^{k-1}-I_{\mu\nu}^{k-1})-\frac{S_{\mu\nu}^{k-1}-S_{\mu\nu}^k}{\Delta\tau_{\mu\nu}^{k-1}}b^{k-1}\bigg]\\
d_\chi^k &=& -\frac{S_{\mu\nu}^{k-1}}{\chi_{\mu\nu}^{k-1}}b^{k-1}+\frac{|z_{k-1}-z_k|}{2\mu}\bigg[e^{-\Delta\tau_{\mu\nu}^{k-1}}(S_{\mu\nu}^{k-1}-I_{\mu\nu}^{k-1})-\frac{S_{\mu\nu}^{k-1}-S_{\mu\nu}^k}{\Delta\tau_{\mu\nu}^{k-1}}b^{k-1}\bigg] \\
c_\eta^k &=& \frac{S_{\mu\nu}^k}{\eta_{\mu\nu}^k}a^{k-1}\\
d_\eta^k &=& \frac{S_{\mu\nu}^{k-1}}{\eta_{\mu\nu}^{k-1}}b^{k-1}
\end{eqnarray}
when $k > 1;$ otherwise, one can set the coefficients to zero.

 \section{Full Jacobian with background scattering terms}
\label{ap:jac}

Let us recall Eq.~\ref{eq:partial_dev_2} with different indices
\diar{\begin{equation}
\bigg(\frac{\partial I_{j\nu}^i}{\partial n_r^\ell}\bigg) = \beta_r^\ell\bigg(\frac{\partial I_{j\nu}^i}{\partial \chi_{j\nu}^\ell}\bigg)+\gamma_r^\ell\bigg(\frac{\partial I_{j\nu}^i}{\partial \eta_{j\nu}^\ell}\bigg)+\sum_p \chi_\textrm{sca}^p\bigg(\frac{\partial J_\nu^p}{\partial n_r^\ell}\bigg)\bigg(\frac{\partial I_{j\nu}^i}{\partial\eta_{j\nu}^p}\bigg)
.\end{equation}}
If we develop the mean intensity term using the angular quadrature scheme with weights $\omega_\mu$, one can write\diar{\begin{equation}
\label{eq:new_indices}
\bigg(\frac{\partial I_{j\nu}^i}{\partial n_r^\ell}\bigg) = \beta_r^\ell\bigg(\frac{\partial I_{j\nu}^i}{\partial \chi_{j\nu}^\ell}\bigg)+\gamma_r^\ell\bigg(\frac{\partial I_{j\nu}^i}{\partial \eta_{j\nu}^\ell}\bigg)+\sum_p \chi_\textrm{sca}^p\bigg(\frac{\partial I_{j\nu}^i}{\partial\eta_{j\nu}^p}\bigg)\sum_q\omega_q \bigg(\frac{\partial I_{q\nu}^p}{\partial n_r^\ell}\bigg)
.\end{equation}}
Going any further requires simplifying the notations to keep as much clarity as possible. Let us define the following quantities
\diar{\begin{eqnarray}
M_{ij} &=& \bigg(\frac{\partial I_{j\nu}^i}{\partial n_r^\ell}\bigg) \\
B_{ij} &=& \beta_r^\ell\bigg(\frac{\partial I_{j\nu}^i}{\partial \chi_{j\nu}^\ell}\bigg)+\gamma_r^\ell\bigg(\frac{\partial I_{j\nu}^i}{\partial \eta_{j\nu}^\ell}\bigg) \\
Q^k_{ij} &=& \chi_\textrm{sca}^i\bigg(\frac{\partial I_{j\nu}^k}{\partial\eta_{j\nu}^i}\bigg)
.\end{eqnarray}}
Here we have omitted the indices \diar{$r,\ell$, and $\nu$}. From this point on, we no longer mention nor write these indices; however, it should be noted that the final solution should be computed for them as well. Equation \ref{eq:new_indices} can now be expressed as\begin{equation}
\label{eq:mat}
M_{ij}-\sum_p Q^i_{pj}\sum_q\omega_q M_{pq} = B_{ij}
.\end{equation}
In this equation, the unknowns one is seeking are the coefficients $M_{ij}$. Equation \ref{eq:mat} can also be expressed as\begin{equation}
\label{eq:vectors}
\vec x_i-\sum_p\langle \vec x_p, \vec \Omega \rangle (\bold{Q}^i)^\intercal\vec e_p = \vec b_i
,\end{equation}
where $\vec x_i = (M_{i1},...,M_{iN_\mu})^\intercal$, $\vec b_i = (B_{i1},...,B_{iN_\mu})^\intercal$, $\vec \Omega = (\omega_1,...,\omega_{N_\mu})^\intercal$, and $\bold{Q}^i$ is the matrix associated with the coefficients $Q^i_{kl}$. The vector $\vec e_p$ is the $p^\textrm{th}$ canonical basis vector. At this point, the unknowns are gathered into the vectors $\vec x_i$. It should be mentioned that if the background scattering is ignored, $\vec x_i = \vec b_i$ and the solution is therefore simple. Otherwise, let us dot product Eq.~\ref{eq:vectors} with $\vec \Omega$ to obtain\begin{equation}
r_i-\sum_p A_{ip} r_p = b_{\Omega,i}
,\end{equation}
where $r_i = \langle \vec x_i, \vec \Omega \rangle$, $b_{\Omega,i} = \langle \vec b_i, \vec \Omega \rangle$,and $A_{ij} = [\bold{Q}^i\vec \Omega]_j$. The equation can also be written using  matrix notations to obtain\begin{equation}
(\bold{I}-\bold{A})\vec r = \vec b_\Omega
,\end{equation}
where $\bold{I}$ is the identity matrix. This is a simple linear system of unknown $\vec r$ for which the solution is\begin{equation}
\vec r = (\bold{I}-\bold{A})^{-1}\vec b_\Omega
\end{equation}
but can be simplified into
\begin{equation}
\vec r \approx (\bold{I}+\bold{A})\vec b_\Omega
\end{equation}
if the background scattering is weak. Introducing this result into Eq. \ref{eq:vectors} gives the final solution
\begin{equation}
\vec x_i = \vec b_i + \sum_p r_p (\bold{Q}^i)^\intercal\vec e_p
,\end{equation}
and if the background scattering is weak,\begin{equation}
\vec x_i \approx \vec b_i + \sum_p b_{\Omega,p} (\bold{Q}^i)^\intercal\vec e_p
.\end{equation}

\end{appendix}

\end{document}